\documentclass[useAMS,usenatbib,onecolumn]{mnras}
\usepackage{amsmath,bm,amssymb}
\usepackage{dcolumn}
\usepackage[utf8]{inputenc}
\usepackage{graphics,graphicx}
\usepackage{float}
\usepackage{threeparttable}
\usepackage{url}
\usepackage{epstopdf}
\usepackage{siunitx}

\usepackage{booktabs}

\usepackage{longtable}

\usepackage{color}

\newcommand{\cm}{cm$^{-1}$}

\newcommand{\X}{$X\,{}^{1}\Sigma^{+}$}
\newcommand{\E}{$E\,{}^{1}\Sigma^{+}$}
\newcommand{\A}{$A\,{}^{1}\Pi$}

\newcommand{\C}{$C\,{}^{1}\Sigma^{-}$}
\newcommand{\D}{$D\,{}^{1}\Delta$}

\newcommand{\astate}{$a\,{}^{3}\Sigma^{+}$}
\newcommand{\bstate}{$b\,{}^{3}\Pi$}
\newcommand{\cstate}{$e\,{}^{3}\Sigma^{-}$}
\newcommand{\dstate}{$d\,{}^{3}\Delta$}

\newcommand{\sio}{$^{28}$Si$^{16}$O}

\newcommand{\name}{SiOUVenIR}

\graphicspath{{figures/}}
\newcommand{\ai}{\textit{ab initio}}

\title[ExoMol line lists -- {XLIV}. SiO]{ExoMol line lists -- {XLIV}. IR and UV line list for silicon monoxide ($^{28}$Si$^{16}$O)}

\date{\today}

\author[ExoMol et al.]{Sergei  N. Yurchenko$^{1}$, Jonathan Tennyson$^{1}$,
Anna-Maree Syme$^{2}$,  Ahmad Y. Adam$^{3}$, \newauthor{Victoria H. J. Clark$^{1}$, Bridgette Cooper$^{1}$, C. Pria Dobney$^{1}$,
Shaun T. E. Donnelly$^{2}$,  }
\newauthor{Maire N. Gorman$^{2}$, Anthony E. Lynas-Gray$^{1,5,6}$, Thomas Meltzer$^{1}$, Alec Owens$^{1}$, Qianwei Qu$^{1}$, }
\newauthor{ Mikhail Semenov$^{1,5}$, Wilfrid Somogyi$^{1}$,  Apoorva Upadhyay$^{1}$, Samuel Wright$^{1}$, Juan C. Zapata Trujillo$^{2}$} \vspace*{4mm}\\
$^1$ Department of Physics and Astronomy, University College London, Gower Street, WC1E 6BT London, UK\\
$^2$ School of Chemistry, University of New South Wales, 2052, Sydney, Australia\\
$^3$ Physikalische und Theoretische Chemie, Fakult{\"a}t f{\"u}r Mathematik und Naturwissenschaften, \\
Bergische Universit{\"a}t Wuppertal, D-42097 Wuppertal, Germany\\
$^4$ Department of Physics, University of Aberystwyth, Ceredigion, Wales SY23 3BZ, UK \\
$^5$ Department of Physics, University of Oxford, Keble Rd., Oxford OX1 3RH, UK\\
$^6$ Department of Physics and Astronomy, University of the Western Cape, Bellville 7535, South Africa\\
$^7$ Department of Science and Research, Moscow Witte University, 2nd Kozhukhovskiy passage,  Moscow, Russian Federation
}

\begin{document}

\label{firstpage}

\maketitle

\pagerange{\pageref{firstpage}--\pageref{lastpage}}

\begin{abstract}

A new silicon monoxide ($^{28}$Si$^{16}$O) line list covering infrared, visible and ultraviolet regions called \name\ is presented. This line list extends the infrared EBJT ExoMol line list by including vibronic transitions to the \A\ and \E\ electronic states. Strong perturbations to the \A\ band system are accurately modelled through the treatment of 6 dark electronic states: \C, \D, \astate, \bstate, \cstate\ and \dstate. Along with the \X\ ground state, these 9 electronic states were used to build a comprehensive spectroscopic model of SiO using a combination of empirical and \ai\ curves, including the potential energy (PE), spin-orbit (SO), electronic angular momentum (EAM) and (transition) dipole moment curves. The  \ai\ PE and coupling curves, computed at the multireference configuration interaction (MRCI) level of theory, were refined by fitting their analytical representations to {2617} experimentally derived SiO energy levels determined from {97} vibronic bands belonging to the $X$--$X$, $E$--$X$ and $A$--$X$ electronic systems through the MARVEL procedure. {112} observed forbidden transitions from the $C$--$X$, $D$--$X$, $e$--$X$, and $d$--$X$ bands were assigned using our predictions, and these could be fed back into the MARVEL procedure. The \name\ line list was computed using published \textit{ab initio} transition dipole moments for the $E$--$X$ and $A$--$X$ bands; %Bauschlicher (2016, Chem. Phys. Lett., 658, 76)
the line list is suitable for temperatures up to 10\,000 K and for wavelengths longer than {140} nm. \name\ is available from \url{www.exomol.com} and the CDS database.

\end{abstract}


\section{Introduction}
\label{sec:intro}

%\verb!XandEandA_Duo_03_current_best_combination_of_exp_and_Duo-X.txt!

Silicon monoxide (SiO) has been observed in a wide variety of astronomical environments since its original detection in the interstellar medium~\citep{71WiPeJe.SiO}, and has been found in late-type stars~\citep{71CuGaKn.SiO} in addition to being a key astrophysical maser~\citep{74SnBuxx.SiO}. Recently, there has been speculation that SiO will be present in the atmospheres of exoplanets, notably in hot rocky planets close to their host star where extreme temperatures cause vaporization of the silicate surface of the planet~\citep{12ScLoFe.SiO}. Significant quantities of SiO can accumulate in a planetary atmosphere and a detection of SiO would shed light on the interaction between the atmosphere and the crust of hot rocky exoplanets at high temperatures~\citep{20HeWoHe}. Calculations have shown that SiO absorption should dominate the infrared (IR) and ultraviolet (UV) wavelength regions in the atmospheres of hot rocky super-Earths with molten surfaces, with prominent features at 4, 10 and 100~$\mu$m~\citep{15ItIkKa.exo}. SiO can also become a major opacity source at wavelengths shorter than 0.35~$\mu$m and dominate the transmission spectra of ultra-hot Jupiters~\citep{20LoFuSi.SiO}.

Any observation of SiO in exoplanets relies on accurate molecular opacity data for this molecule. Currently, models use the SiO rotation-vibration-electronic (rovibronic) line list of \citet{11Kurucz.db}, which covers the \X--\X, \A--\X\ and \E--\X\ electronic bands. Computational approaches to construct molecular line lists have undergone considerable development since the Kurucz SiO line list~\citep{jt654,jt626}. Notably, a robust computer program \textsc{Duo}~\citep{jt609} has been developed to compute the rovibronic spectra of diatomic molecules and is able to treat multiple interacting electronic states and their couplings, as is the case in SiO. Secondly, an efficient algorithm known as MARVEL (Measured Active Rotational-Vibrational Energy Levels)~\citep{jt412,07CsCzFu.method,12FuCsxx.methods,jt750} can analyse all the available published spectroscopic data (line positions) on a molecule and convert it into a consistent set of highly accurate empirical-quality energy levels. These can be used to empirically refine the spectroscopic model of a molecule, ultimately yielding line positions that can meet the demands of high-resolution spectroscopy of exoplanets~\citep{18Birkby-RV} and be incorporated into the line list directly by replacing calculated values. We thus find it worthwhile to generate a new SiO rovibronic line list that exploits these tools within the ExoMol computational framework~\citep{jt731}. The line list has been produced for the ExoMol Database~\citep{jt528,jt631,jt810}, which is providing key molecular data on a range of important molecules for the study of exoplanets and other hot atmospheres.

Experimental sources of SiO spectroscopic data selected for this work include those of \citet{68Torring.SiO,71ElLaxx.SiO,76FiLaRe.SiO,76BrReCo.SiO,77MaClDe.SiO,81LoMaOl.SiO,91MoGoVr.SiO,92WaLixx.db,94TsOhHi.SiO,95CaKlDu.SiO,96WaLiHi.db,03SaMcTh.SiO} and \citet{13MuSpBi.SiO}.
Experimentally, the \A\ -- \X\ band of  SiO has  been extensively studied in the 1970s by  \citet{76FiLaRe.SiO} and \citet{76BrReCo.SiO}. This band undergoes strong perturbations with nearby `dark' electronic states \C, \D, \astate, \bstate, \cstate\ and \dstate, i.e. states with transitions forbidden from the \X\ state. These perturbations were thoroughly analysed by \citet{76FiLaRe.SiO} and \citet{76BrReCo.SiO} with a significant number of transitions to dark states characterised. This valuable information plays an important role in our calculations as it allows us to use experimental  constraints for these 6 dark states, which are otherwise not available.

The strong \E\ -- \X\ system of \sio\ was studied by \citet{71ElLaxx.SiO} who reported rovibronic transitions up to vibrational levels with $v'=14$. For this system, the perturbations were not reported with the necessary level of detail and are therefore not covered in our work. The $F$--$X$, $G$--$X$, $I$--$X$, $K$--$X$, $J$--$X$, $M$--$X$, $L$--$X$, $N$--$X$, $P$--$X$, $O$--$X$ band systems have also been observed experimentally \citep{73LaReEl.SiO}, but not at a resolution high enough for our analysis. These data were not included in the present study.

Lifetimes of SiO states were studied experimentally by \citet{72SmLixx.SiO, 73ElSmxx.SiO} and comparison with these data provides an important test of the transition intensities predicted by our model.

Some of the key  \ai\ studies of SiO are by \citet{93LaBaxx.SiO,98DrSpEd.SiO,03ChChD1.SiO,08XuLuLi.SiO,12ShLiSu.SiO,16Bauschlicher.SiO,19FeZhxx.SiO} and \citet{19FeZhxxi.SiO}.
The most valuable is the recent work by \citet{16Bauschlicher.SiO} who reported high level potential energy and transition dipole moment curves of SiO, which we have made extensive use of in this work.

We present a new $^{28}$Si$^{16}$O line list, called \name,  built using an accurate spectroscopic model: empirical potential energy curves (PECs), spin-orbit curves, electronic angular momentum curves (EAMCs) and \ai\  dipole moment curves (DMCs) and  transition dipole moment curves (TDMCs). As part of the construction of the empirical curves, a MARVEL (Measured Active Rotational-Vibrational Energy Levels) set of empirical energies for $^{28}$Si$^{16}$O are  produced. The line list covers  the \X--\X, \A--\X\ and \E--\X\ electronic systems and includes transitions from  the `dark' bands.  The \name\ line list is suitable for temperatures up to 10\,000 K and spans wavelengths longer than 140 nm (72\,000 \cm)
%i.e. up to the dissociation of the $X$ state {correct?}.
The \name\ line list builds upon the previous ExoMol study of \citet{jt563}, which produced the EBJT rotation-vibration line lists for the ground electronic state $X\,^1\Sigma^+$ of the main isotopologue $^{28}$Si$^{16}$O and the four monosubstituted isotopologues $^{29}$Si$^{16}$O, $^{30}$Si$^{16}$O, $^{28}$Si$^{18}$O, and $^{28}$Si$^{17}$O.

This paper is structured as follows:  In Section~\ref{sec:marvel}, we perform an analysis of all the published spectroscopic literature on SiO using the MARVEL procedure. The underlying potential energy curves (PECs), dipole moment curves (DMCs), and angular momentum coupling curves of our SiO spectroscopic model are discussed in Section~\ref{sec:pec}. The solution of the coupled Schr\"{o}dinger equations using \textsc{Duo}~\citep{jt609}, fitting of the curves via deperturbation of states and the production of the  rovibronic line list  are discussed in Section \ref{sec:duo}. The line list and its applications, including lifetimes and comparisons with lab and stellar spectra are presented in Section~\ref{s:linelist}. We conclude in Section~\ref{sec:conc}.

\section{MARVEL}
\label{sec:marvel}

All available experimental transition frequencies of SiO were extracted from the published spectroscopic literature and analysed using the MARVEL  algorithm~\citep{jt412,07CsCzFu.method,12FuCsxx.methods,jt750}. This procedure takes a set of assigned transition frequencies with measurement uncertainties and converts it into a consistent set of empirical-quality energy levels with the uncertainties propagated from the input transitions. The SiO data extracted from the literature covers the three main bands involving the $X\,^1\Sigma^+$, $A\,^1\Pi$, and $E\,^1\Sigma^+$ electronic states: \X--\X, \A--\X\ and \E--\X\, as summarised in Table~\ref{tab:trans}. Due to interactions with other electronic states, 112 of the measured rovibronic transitions belong to the four forbidden systems     \C--\X, \D--\X, \cstate-\X\ and \dstate-\X, which are usually not identified in the experimental data but can be assigned using our variational calculations. It is important for MARVEL that all transitions are interconnected into a single network, which is not always the case. Here we include the empirical energies of SiO from the accurate ExoMol EBJT line list \citep{jt563} for $v\le {7}$ and $J\le 100$ in order to fill  gaps and connect all experimental data into  a single network. According to our experience,  EBJT calculated energies are accurate enough for this task.
%\red{JT: we cannot call the EBJT levels Duo! They were computed with Level. Call them either EBJT or probably better 13BaYuTe. This will need changing in the MARVEL files.}

\subsection{Description of experimental sources}

\begin{figure}
   % \centering
    \includegraphics[width = 0.8\textwidth]{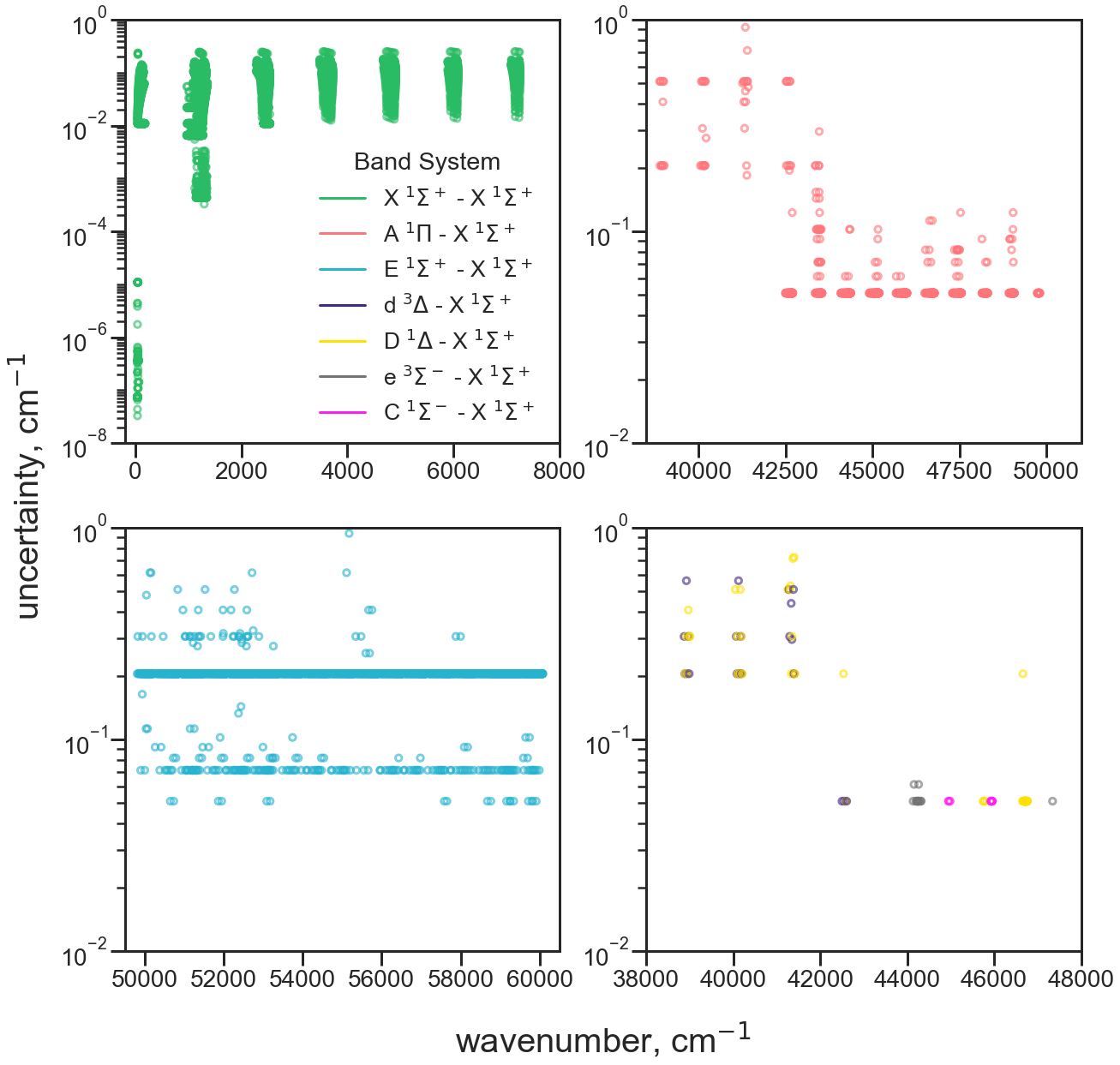}
    \caption{Spread of uncertainties across transition wavenumber for each electronic band within SiO. Forbidden bands are grouped together in the bottom right plot. Note the different y axis for the \X--\X subplot (top left panel). }
    \label{m:bands}
\end{figure}

\begin{table*}
    \centering
    \caption{Breakdown of the assigned transitions by electronic bands for the sources used in this study. \textit{A} is the number of available transitions and \textit{V} is the number of validated. The mean and maximum uncertainties obtained using the MARVEL procedure are given in \cm.   }
    \label{tab:trans}
    \resizebox{\textwidth}{!}{%
    \begin{tabular}{p{2.4cm}p{10cm}p{1.4cm}p{2.4cm}p{2.6cm}p{1.6cm}}
          \toprule
Electronic Band & Vibrational Bands & $J$ Range & Wavenumber Range \cm{} & Mean/Max    &  A/V    \\ \midrule
 \textbf{13BaYuTe} \\
X $^1\Sigma^+$-X $^1\Sigma^+$  & (0-0), (1-0), (1-1), (2-0), (2-1), (2-2), (3-0), (3-1), (3-2), (3-3), (4-0), (4-1), (4-2), (4-3), (4-4), (5-0), (5-1), (5-2), (5-3), (5-4), (5-5), (6-0), (6-1), (6-2), (6-3), (6-4), (6-5), (7-1), (7-2), (7-3), (7-4),   & 0 - 103 & 1.4 - 7216.6   & 7.37e-02/2.28e-01   & 3426/3426 \\
\\ \textbf{68Torring} \\
X $^1\Sigma^+$-X $^1\Sigma^+$  & (0-0), (1-1), (2-2), (3-3),   & 0 - 1 & 1.4 - 1.4   & 3.27e-06/4.00e-06   & 4/4 \\
\\ \textbf{71ElLa} \\
E $^1\Sigma^+$-X $^1\Sigma^+$  & (0-0), (0-1), (0-2), (1-0), (1-1), (1-2), (10-0), (11-0), (12-0), (13-1), (14-1), (2-0), (2-1), (2-2), (3-0), (3-1), (4-0), (5-0), (6-0), (7-0), (8-0), (9-0),   & 2 - 64 & 49762.1 - 60025.8   & 1.91e-01/9.20e-01   & 2079/2076 \\
\\ \textbf{76FiLaRe} \\
A $^1\Pi$-X $^1\Sigma^+$  & (0-0), (1-0), (2-0), (3-0), (4-0), (5-0), (6-0), (7-0), (8-0), (9-0),   & 0 - 49 & 42439.1 - 49752.3   & 5.60e-02/5.00e-01   & 1032/1032 \\
D $^1\Delta$-X $^1\Sigma^+$  & (6-0),   & 35 - 36 & 42453.1 - 42558.4   & 5.00e-02/5.00e-02   & 3/3 \\
d $^3\Delta$-X $^1\Sigma^+$  & (14-0), (15-0), (9-0),   & 11 - 41 & 42487.9 - 46720.6   & 5.88e-02/2.00e-01   & 34/34 \\
e $^3\Sigma^-$-X $^1\Sigma^+$  & (14-0), (15-0), (7-0), (9-0),   & 9 - 38 & 44089.1 - 48255.8   & 5.08e-02/6.00e-02   & 24/24 \\
C $^1\Sigma^-$-X $^1\Sigma^+$  & (10-0), (11-0),   & 13 - 42 & 44898.4 - 45912.8   & 5.00e-02/5.00e-02   & 9/9 \\
\\ \textbf{76BrReCo} \\
D $^1\Delta$-X $^1\Sigma^+$  & (6-1), (6-2), (6-3),   & 34 - 36 & 38821.0 - 41349.4   & 3.18e-01/5.50e-01   & 18/18 \\
A $^1\Pi$-X $^1\Sigma^+$  & (0-1), (0-2), (0-3),   & 28 - 39 & 38825.2 - 41372.0   & 3.82e-01/9.00e-01   & 42/42 \\
d $^3\Delta$-X $^1\Sigma^+$  & (9-1), (9-2), (9-3),   & 28 - 39 & 38841.4 - 41374.5   & 3.39e-01/7.10e-01   & 24/24 \\
\\ \textbf{77MaClDe} \\
X $^1\Sigma^+$-X $^1\Sigma^+$  & (0-0), (1-1), (2-2), (3-3), (4-4),   & 1 - 7 & 2.9 - 10.1   & 1.00e-05/1.00e-05   & 17/17 \\
\\ \textbf{81LoMaOI} \\
X $^1\Sigma^+$-X $^1\Sigma^+$  & (1-0), (2-1), (3-2), (4-3), (5-4),   & 0 - 61 & 1213.0 - 1294.0   & 1.48e-03/3.00e-03   & 35/35 \\
\\ \textbf{91MoGoVr} \\
X $^1\Sigma^+$-X $^1\Sigma^+$  & (1-1), (10-10), (11-11), (13-13), (14-14), (15-15), (16-16), (17-17), (18-18), (19-19), (2-2), (20-20), (21-21), (22-22), (25-25), (3-3), (30-30), (31-31), (32-32), (33-33), (34-34), (36-36), (37-37), (38-38), (39-39), (4-4), (40-40), (5-5), (6-6), (7-7), (8-8), (9-9),   & 5 - 9 & 7.6 - 11.4   & 3.56e-07/5.67e-07   & 62/62 \\
\\ \textbf{92WaLi} \\
X $^1\Sigma^+$-X $^1\Sigma^+$  & (2-0), (3-1), (4-2), (5-3), (6-4),   & 3 - 101 & 2381.4 - 2497.3   & 1.01e-02/2.00e-02   & 507/507 \\
\\ \textbf{94TsOhHi} \\
X $^1\Sigma^+$-X $^1\Sigma^+$  & (2-0), (3-1), (4-2), (5-3),   & 2 - 120 & 2420.3 - 2486.6   & 1.10e-02/2.00e-02   & 50/50 \\
\\ \textbf{95CaKlDu$^a$} \\
X $^1\Sigma^+$-X $^1\Sigma^+$  & (1-0), (10-9), (101-100), (102-101), (103-102), (104-103), (105-104), (106-105), (11-10), (12-11), (13-12), (2-1), (3-2), (4-3), (5-4), (6-5), (61-60), (62-61), (63-62), (64-63), (65-64), (66-65), (7-6), (8-7), (9-8),   & 0 - 140 & 926.6 - 1241.3   & 9.52e-03/5.00e-02   & 1661/1622 \\
\\ \textbf{95CaKlDu:lab} \\
X $^1\Sigma^+$-X $^1\Sigma^+$  & (1-0), (2-1), (3-2), (4-3), (5-4),   & 1 - 82 & 1080.9 - 1311.9   & 5.54e-04/6.30e-03   & 375/375 \\
\\ \textbf{03SaMcTh} \\
X $^1\Sigma^+$-X $^1\Sigma^+$  & (0-0), (1-1), (10-10), (11-11), (12-12), (13-13), (14-14), (15-15), (16-16), (17-17), (18-18), (19-19), (2-2), (20-20), (21-21), (22-22), (23-23), (24-24), (25-25), (26-26), (27-27), (28-28), (29-29), (3-3), (30-30), (31-31), (32-32), (33-33), (34-34), (35-35), (36-36), (37-37), (38-38), (39-39), (4-4), (40-40), (41-41), (42-42), (43-43), (44-44), (45-45), (5-5), (6-6), (7-7), (8-8), (9-9),   & 0 - 1 & 1.0 - 1.4   & 1.02e-07/6.00e-07   & 46/46 \\
\\ \textbf{13MuSpBi} \\
X $^1\Sigma^+$-X $^1\Sigma^+$  & (0-0), (1-1), (2-2), (3-3), (4-4), (46-46), (47-47), (48-48), (49-49), (5-5), (50-50), (51-51),   & 0 - 19 & 0.9 - 27.5   & 1.77e-07/5.00e-07   & 29/29 \\
   \bottomrule
   \end{tabular}
   }
{\flushleft
$^a$ From \citet{13MuSpBi.SiO}. \\ }
\end{table*}

Experimental transitions from the ground state system $X$--$X$ are the same as used by \citet{jt563} to construct the EBJT line list. The other sources, denoted by the MARVEL tag, are follows:

\textbf{13MuSpBi} \citep{13MuSpBi.SiO}: reported 29 rotational microwave transitions of the $X$--$X$ system for a number of isotopologues of SiO, including \sio, for vibrational levels with $v$ up to 51. They have also compiled an extensive source of all important spectroscopic data for $X$--$X$ of SiO from other sources, some of which were sourced to the original lab or observational data, for example their version of the sunspot IR line positions by \citet{95CaKlDu.SiO} which appears to give more decimal places.

\textbf{68Torring}: Four microwave transitions by \citet{68Torring.SiO}.

\textbf{77MaClDe}: 17 microwave transitions by \citet{77MaClDe.SiO}.

\textbf{81LoMaOl}: 35 lines in the  1240 \cm\ region by \citet{81LoMaOl.SiO}.

\textbf{91MoGoVr}: Microwave data (62 lines) by \citet{91MoGoVr.SiO} taken from the compilation by \citet{13MuSpBi.SiO}.

\textbf{94TsOhHi}: IR data (50 lines) by \citet{94TsOhHi.SiO} from the region of the first overtone, 2400~\cm, observed in spectra of giant M-stars.

\textbf{95CaKlDu:lab} and \textbf{95CaKlDu:sunspot}: IR sunspot spectra of SiO by \citet{95CaKlDu.SiO}, which were then recompiled by \citet{13MuSpBi.SiO}.  Here we use the data provided by \citet{13MuSpBi.SiO}, which appeared in two parts, as lab and sunspot data. We retain this structure in our analysis.

%\textbf{98ChSaxx} \citep{}: 10 microwave lines for 28Si18O  by \citet{98ChSaxx.SiO}.

\textbf{92WaLi}: 507 IR line positions covering the region of the 1st overtone band (2400~\cm) of \sio\ were extracted from the  IR Spectral  Atlases of the Sun by \citet{92WaLixx.db}, see  \citet{96WaLiHi.db}.
There is a further sunspot Atlas due to \citet{94WaLiBe.db} which contains identified SiO lines. However this Atlas does not provide line positions so it was not used.

\textbf{03SaMcTh}: 46 rotational (microwave) transitions of SiO by \citet{03SaMcTh.SiO} covering high vibrational excitations, up to $v=44$.

\textbf{76FiLaRe,76BrReCo} \citep{76FiLaRe.SiO,76BrReCo.SiO}: The \A\ -- \X\ system was reported by two main sources, \citet{76FiLaRe.SiO}, (0-0), (1-0), (2-0), (3-0), (4-0), (5-0), (6-0), (7-0), (8-0), (9-0), 1102 lines,  and \citet{76BrReCo.SiO} (0-1), (0-2), (0-3), 84 lines. This system is heavily perturbed by  \C, \D, \astate, \bstate, \cstate\ and \dstate.  In this study, a significant number of forbidden transitions to these dark states were measured, which were all assigned as additional $A$--$X$ transitions. MARVEL requires that all transitions are uniquely identified. To this end we initially gave a dummy label $A'$ to all duplicates, which were then properly resolved by comparing to the calculated values using \textsc{Duo}  (see below). These data are very important as they provide (the only)  access to these forbidden states, which strongly affect the behavior of the `bright' states.

\textbf{71ElLa}: UV data covering the $E$--$X$ system by \citet{71ElLaxx.SiO}, 2079 lines, $v'_{\rm max} =14$.

\textbf{13BaYuTe}: 3426 pseudo-experimental values for $v=0\ldots 7$ and $J=0\ldots 100$ constructed from the EBJT energies as `transitions' from the zero-point state, \X, $v=0$, $J=0$ to fill gaps in the MARVEL set and connect otherwise disjoint clusters (so called `orphans' \citet{11CsFuxx.marvel}). They play the same role as MARVEL's `Magic numbers'. The uncertainties $\sigma$ of these pseudo-experimental energies were decided based on the following scheme:
$$
\sigma =
\left\{
\begin{array}{cc}
    0.01 + 0.0005 J(J+1) & v\le 6,  \\
    0.2 + 0.0005 J(J+1) &  v = 7
\end{array}
\right.
$$

The vibrational analysis of the higher $E$--$X$, $G$--$X$ band systems by \citet{54BaRoxx.SiO} was not included in the current study as this work does not contain line positions.

% \begin{figure}
%     \centering
%     \includegraphics[width = 0.5\textwidth]{figures/MARVEL_Sourcevsunc.png}
%     \caption{The spread of the uncertainties of the transitions taken from each data source.}
%     \label{m:sources}
% \end{figure}

\begin{figure}
    \centering
    \includegraphics[width=0.75\textwidth]{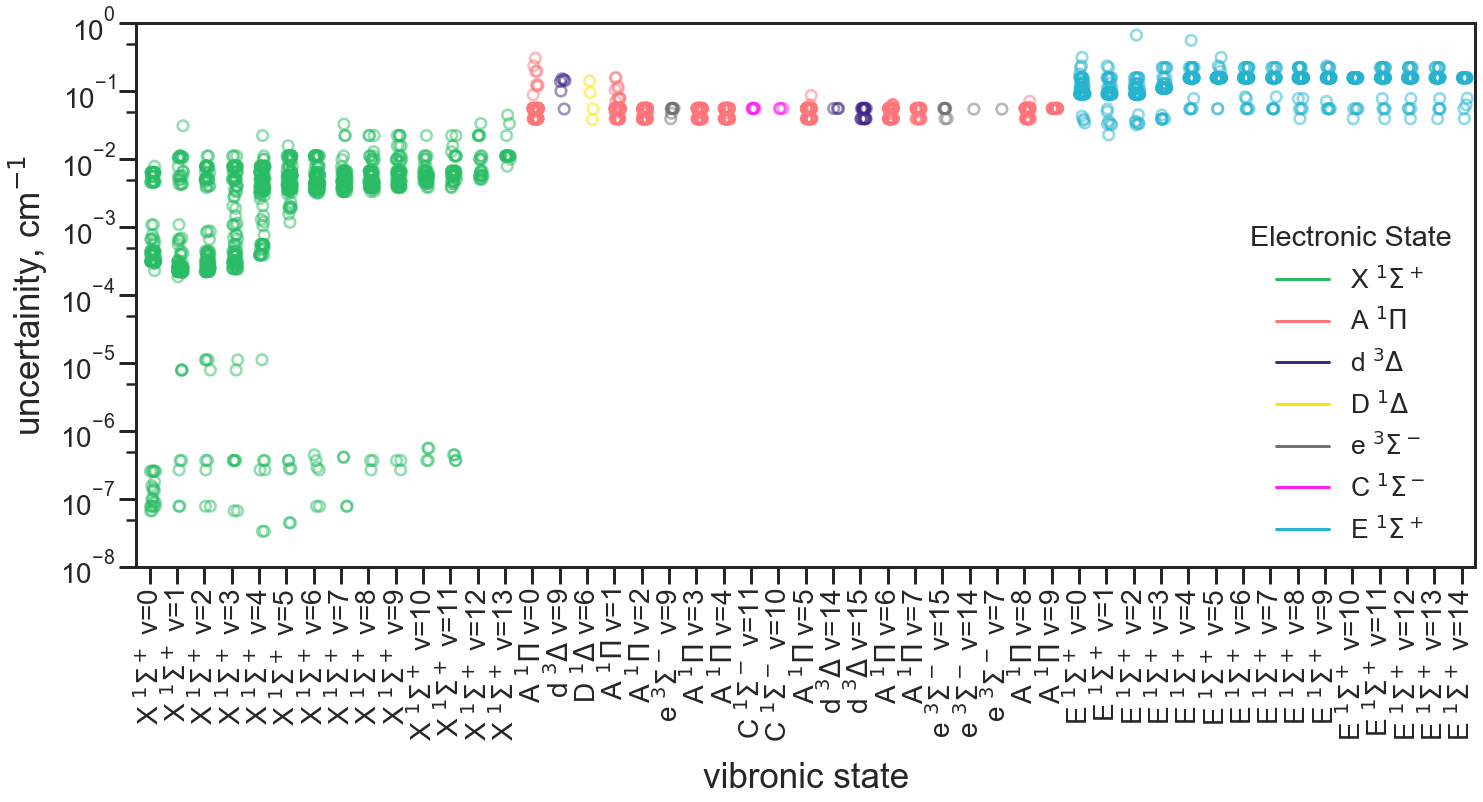}
    \caption{The spread of uncertainties for each vibronic state.}
    \label{m:vibronic}
\end{figure}

\begin{figure}
    \centering
    \includegraphics[width= 0.5\textwidth]{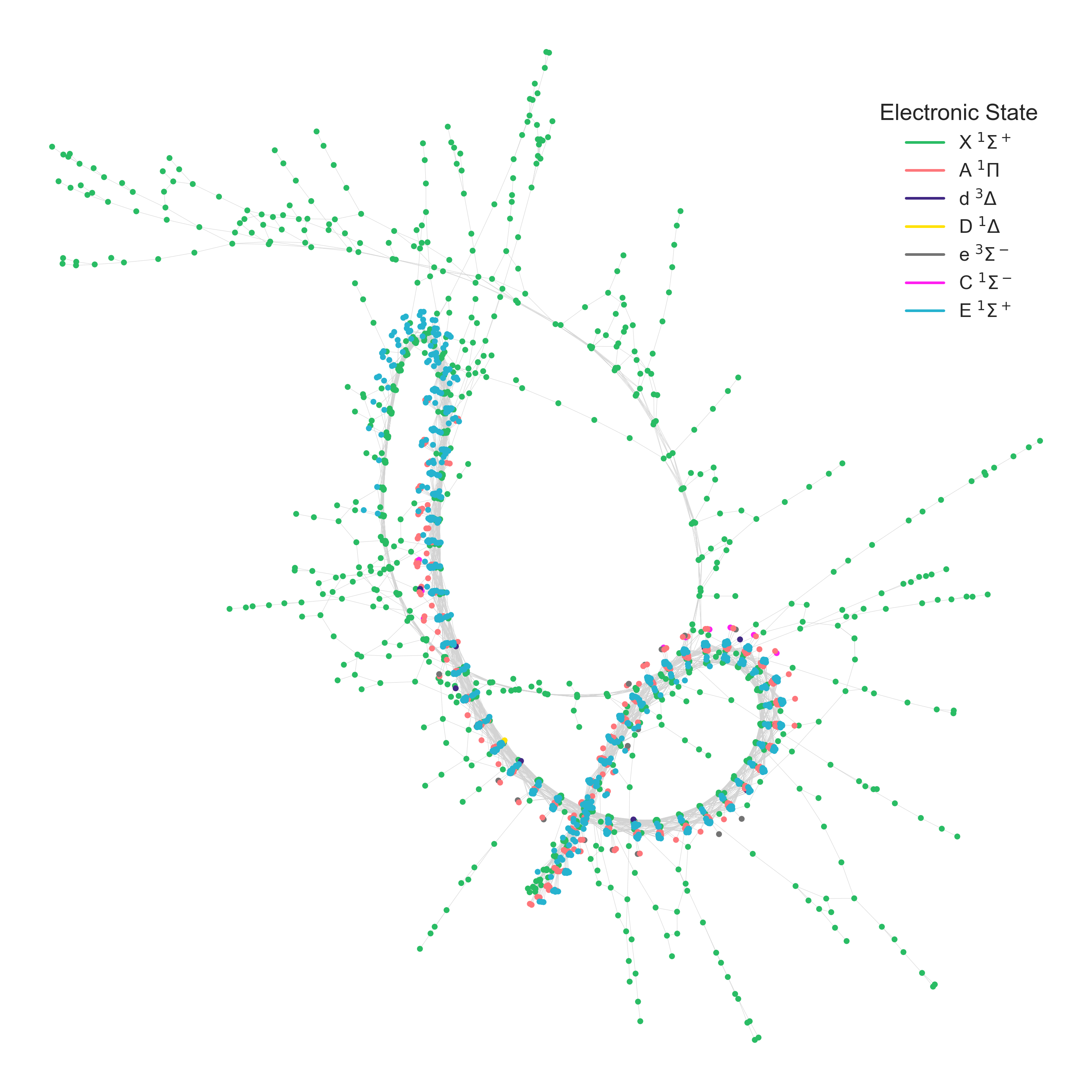}
    \caption{Rovibrational spectroscopic networks for SiO. Each node represents a unique rovibronic energy level with the color indicating the electronic state and the connections between nodes are from input transition data.}
    \label{m:SN}
\end{figure}

In total, {6051} experimental and 3426 pseudo-experimental (13BaYuTe) transitions were processed via the online MARVEL app (available through a user-friendly web interface at \href{http://kkrk.chem.elte.hu/marvelonline}{http://kkrk.chem.elte.hu/marvelonline}) using the Cholesky (analytic) approach with a {0.5}~cm$^{-1}$ threshold on the uncertainty of the ``very bad'' lines. The spread of transition wavenumbers and their uncertainties is shown in Fig.~\ref{m:bands}, split by the electronic bands of the transitions. Unsurprisingly the \X{}--\X{} band  has the lowest uncertainties, reaching an uncertainty of $10^{-7}$~\cm\ for rotational transitions from \citet{03SaMcTh.SiO} and \citet{13MuSpBi.SiO}. The final MARVEL process for SiO resulted in one main spectroscopic network, shown in Fig.~\ref{m:SN}, containing 2617 energy levels with rotational excitation up to $J=103$ for molecular states below 61\,881~cm$^{-1}$. The energy levels are described in Table \ref{tab:el}, giving the $J$ and energy range for each vibronic state, as well as the sources of data that contribute to these energy levels. The spread of uncertainties for each vibronic state is shown in Fig.~\ref{m:vibronic}, clearly showing the higher accuracy and reliability of the \X{} state energy levels. These energy levels were used in \textsc{Duo} to refine our rovibronic spectroscopic model (PECs, SOCs and EAMCs) corresponding to all states but \X, which was kept unchanged. We also did not include the EAM coupling into the fit between $A$ and $X$ in order not to destroy the integrity of the $X$ PEC. The MARVEL input transitions and output energy files are given as part of the supplementary material. Figure~\ref{m:SN} is an illustration of the MARVEL spectroscopic network where the nodes show the ro-vibrationic energy levels of each electronic state and the lines between them represent the transitions. Almost all energy levels from higher electronic states transition to low vibrational states from \X{} which leads to the many offshoots of higher ro-vibrational levels of the \X{} state.
% \red{Alec: Any explanation of the MARVEL spectroscopic network figure?}.

\begin{table*}
\centering
\caption{Summary of  experimentally-derived MARVEL energy levels,
    including uncertainties and data sources in SiO. \textit{No} is the number of energy levels in that vibronic state; \textit{Mean/Max} are the mean and maximal values of uncertainties. \label{tab:el}}
\resizebox{\textwidth}{!}{%
\begin{tabular}{p{1cm}p{1.6cm}p{2.8cm}p{0.6cm}p{2.0cm}p{12cm}}
\toprule
$v$ & $J$ range & E-range (\cm) & No &  Mean/Max & Sources \\ \midrule
\textbf{X $^1\Sigma^+$} \\
0 & 0 - 103 & 0.0 - 7643.7 & 104 & 0.001/0.007 & 03SaMcTh, 13MuSpBi, 13BaYuTe, 68Torring, 77MaClDe, 71ElLa, 76FiLaRe, 81LoMaOI, 92WaLi, 94TsOhHi, 95CaKlDu:lab, 95CaKlDu:sunspot \\
1 & 0 - 102 & 1229.6 - 8675.5 & 102 & 0.002/0.028 & 03SaMcTh, 13MuSpBi, 13BaYuTe, 68Torring, 77MaClDe, 81LoMaOI, 91MoGoVr, 95CaKlDu:lab, 95CaKlDu:sunspot, 71ElLa, 76BrReCo, 92WaLi, 94TsOhHi \\
2 & 0 - 103 & 2447.3 - 9983.1 & 104 & 0.002/0.01 & 03SaMcTh, 13MuSpBi, 13BaYuTe, 68Torring, 77MaClDe, 81LoMaOI, 91MoGoVr, 92WaLi, 94TsOhHi, 95CaKlDu:lab, 95CaKlDu:sunspot, 71ElLa, 76BrReCo \\
3 & 0 - 103 & 3653.2 - 11135.1 & 102 & 0.003/0.01 & 03SaMcTh, 13MuSpBi, 13BaYuTe, 68Torring, 77MaClDe, 81LoMaOI, 91MoGoVr, 92WaLi, 94TsOhHi, 95CaKlDu:lab, 95CaKlDu:sunspot, 76BrReCo \\
4 & 0 - 103 & 4847.2 - 12275.3 & 104 & 0.004/0.02 & 03SaMcTh, 13MuSpBi, 13BaYuTe, 77MaClDe, 81LoMaOI, 91MoGoVr, 92WaLi, 94TsOhHi, 95CaKlDu:lab, 95CaKlDu:sunspot \\
5 & 0 - 102 & 6029.5 - 13264.1 & 103 & 0.005/0.014 & 03SaMcTh, 13MuSpBi, 13BaYuTe, 81LoMaOI, 91MoGoVr, 92WaLi, 94TsOhHi, 95CaKlDu:lab, 95CaKlDu:sunspot \\
6 & 0 - 101 & 7200.0 - 14244.5 & 102 & 0.005/0.01 & 03SaMcTh, 13BaYuTe, 91MoGoVr, 92WaLi, 95CaKlDu:sunspot \\
7 & 0 - 100 & 8358.8 - 15216.6 & 82 & 0.005/0.03 & 03SaMcTh, 13BaYuTe, 91MoGoVr, 95CaKlDu:sunspot \\
8 & 3 - 99 & 9514.1 - 16180.3 & 70 & 0.006/0.02 & 03SaMcTh, 91MoGoVr, 95CaKlDu:sunspot \\
9 & 4 - 98 & 10654.9 - 17135.6 & 60 & 0.007/0.02 & 03SaMcTh, 91MoGoVr, 95CaKlDu:sunspot \\
10 & 5 - 91 & 11785.4 - 17338.4 & 47 & 0.005/0.02 & 03SaMcTh, 91MoGoVr, 95CaKlDu:sunspot \\
11 & 5 - 83 & 12897.5 - 17493.7 & 41 & 0.006/0.02 & 03SaMcTh, 91MoGoVr, 95CaKlDu:sunspot \\
12 & 8 - 80 & 14026.0 - 18239.5 & 22 & 0.01/0.03 & 03SaMcTh, 95CaKlDu:sunspot \\
13 & 9 - 60 & 15126.7 - 17466.3 & 16 & 0.014/0.04 & 03SaMcTh, 91MoGoVr, 95CaKlDu:sunspot \\
\textbf{A $^1\Pi$} \\
0 & 2 - 47 & 42643.8 - 44047.9 & 85 & 0.054/0.275 & 76BrReCo, 76FiLaRe \\
1 & 2 - 45 & 43483.9 - 44759.0 & 85 & 0.052/0.141 & 76FiLaRe \\
2 & 1 - 49 & 44306.4 - 45803.1 & 92 & 0.043/0.05 & 76FiLaRe \\
3 & 1 - 47 & 45124.0 - 46485.8 & 92 & 0.044/0.05 & 76FiLaRe \\
4 & 1 - 49 & 45925.8 - 47388.4 & 92 & 0.043/0.05 & 76FiLaRe \\
5 & 1 - 43 & 46715.6 - 47834.2 & 76 & 0.044/0.078 & 76FiLaRe \\
6 & 1 - 42 & 47494.4 - 48551.2 & 79 & 0.045/0.057 & 76FiLaRe \\
7 & 3 - 34 & 48264.5 - 48949.5 & 45 & 0.046/0.05 & 76FiLaRe \\
8 & 3 - 31 & 49021.0 - 49583.2 & 55 & 0.045/0.064 & 76FiLaRe \\
9 & 4 - 21 & 49766.3 - 50012.0 & 26 & 0.05/0.05 & 76FiLaRe \\
\textbf{C $^1\Sigma^-$} \\
10 & 40 - 42 & 46125.4 - 46203.1 & 3 & 0.05/0.05 & 76FiLaRe \\
11 & 13 - 18 & 46044.6 - 46122.6 & 6 & 0.05/0.05 & 76FiLaRe \\
\textbf{D $^1\Delta$} \\
6 & 35 - 36 & 43433.3 - 43469.6 & 4 & 0.074/0.127 & 76BrReCo, 76FiLaRe \\
\textbf{E $^1\Sigma^+$} \\
0 & 2 - 55 & 52581.6 - 54241.7 & 54 & 0.104/0.283 & 71ElLa \\
1 & 3 - 64 & 53252.4 - 55463.2 & 62 & 0.089/0.212 & 71ElLa \\
2 & 2 - 59 & 53907.8 - 55775.0 & 58 & 0.097/0.6 & 71ElLa \\
3 & 3 - 61 & 54561.1 - 56531.4 & 59 & 0.105/0.2 & 71ElLa \\
4 & 1 - 59 & 55198.0 - 57029.0 & 59 & 0.149/0.503 & 71ElLa \\
5 & 4 - 58 & 55840.5 - 57583.7 & 55 & 0.144/0.283 & 71ElLa \\
6 & 2 - 56 & 56458.2 - 58075.2 & 55 & 0.14/0.2 & 71ElLa \\
7 & 5 - 55 & 57087.1 - 58619.4 & 51 & 0.137/0.2 & 71ElLa \\
8 & 3 - 53 & 57687.6 - 59104.2 & 51 & 0.139/0.2 & 71ElLa \\
9 & 1 - 53 & 58281.3 - 59689.4 & 53 & 0.137/0.212 & 71ElLa \\
10 & 4 - 56 & 58881.7 - 60425.5 & 53 & 0.132/0.141 & 71ElLa \\
11 & 1 - 51 & 59456.0 - 60734.1 & 51 & 0.142/0.2 & 71ElLa \\
12 & 6 - 50 & 60049.3 - 61245.7 & 45 & 0.139/0.2 & 71ElLa \\
13 & 10 - 47 & 60647.2 - 61660.3 & 38 & 0.139/0.2 & 71ElLa \\
14 & 15 - 39 & 61263.9 - 61881.1 & 25 & 0.127/0.141 & 71ElLa \\
\textbf{d $^3\Delta$} \\
9 & 29 - 39 & 43187.3 - 43615.4 & 6 & 0.11/0.136 & 76BrReCo, 76FiLaRe \\
14 & 37 - 41 & 46770.1 - 46957.8 & 3 & 0.05/0.05 & 76FiLaRe \\
15 & 12 - 24 & 46816.1 - 47073.3 & 20 & 0.044/0.05 & 76FiLaRe \\
\textbf{e $^3\Sigma^-$} \\
7 & 15 - 15 & 48403.6 - 48403.6 & 1 & 0.05/0.05 & 76FiLaRe \\
9 & 26 - 37 & 44743.3 - 45160.3 & 10 & 0.048/0.05 & 76FiLaRe \\
14 & 38 - 38 & 48367.7 - 48367.7 & 1 & 0.05/0.05 & 76FiLaRe \\
15 & 9 - 16 & 48319.4 - 48421.0 & 8 & 0.046/0.05 & 76FiLaRe \\
\bottomrule
\end{tabular}
}
\end{table*}

\section{Spectroscopic Model of S\lowercase{i}O}
\label{sec:pec}

\subsection{Potential energy, spin orbit and electronic angular momentum curves}

\begin{figure}
    \centering
    \includegraphics[width=0.80\textwidth]{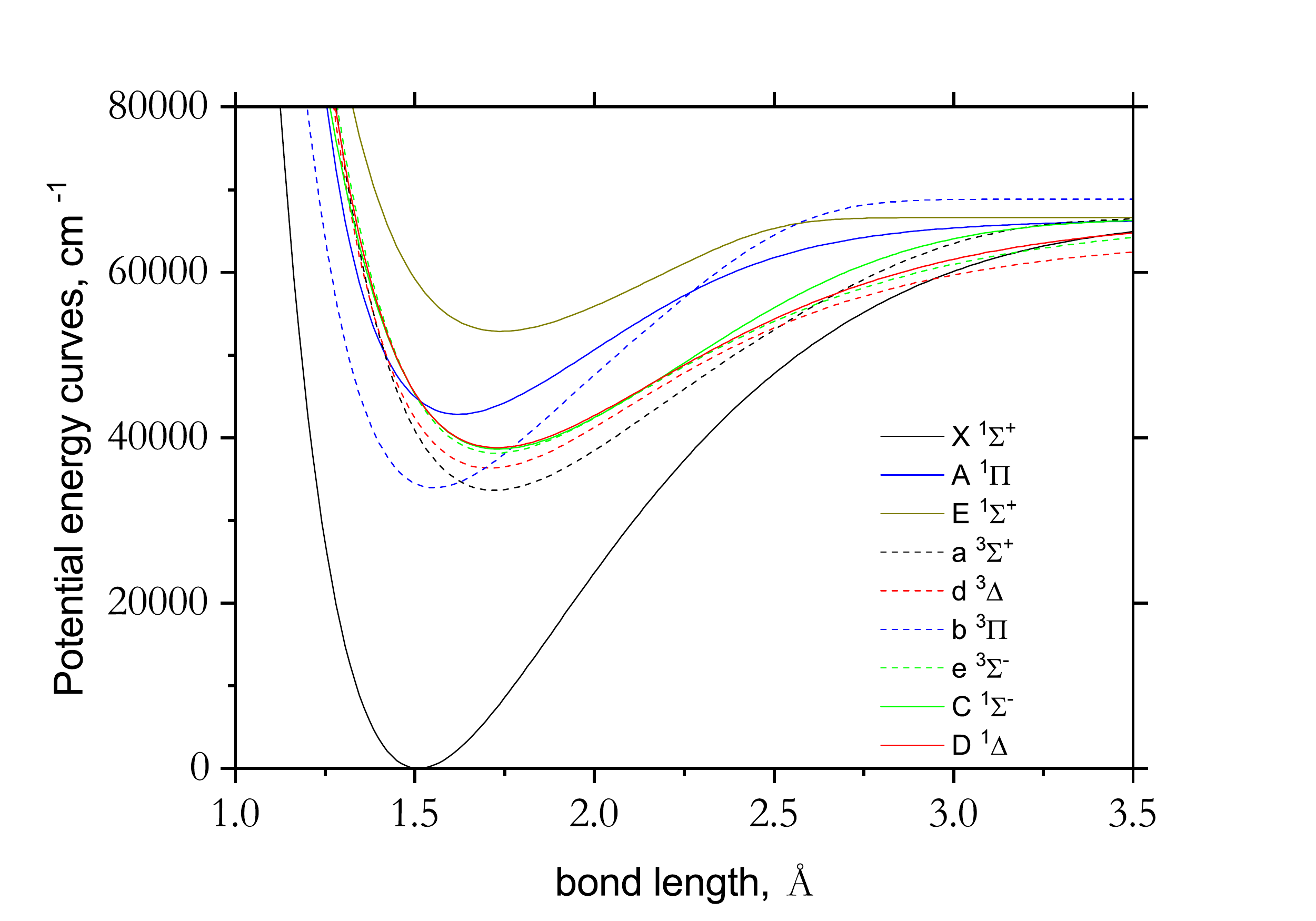}
    \caption{Refined potential energy curves of SiO.}
    \label{fig:PECs}
\end{figure}

\begin{figure}
    \centering
    \includegraphics[width=0.44\textwidth]{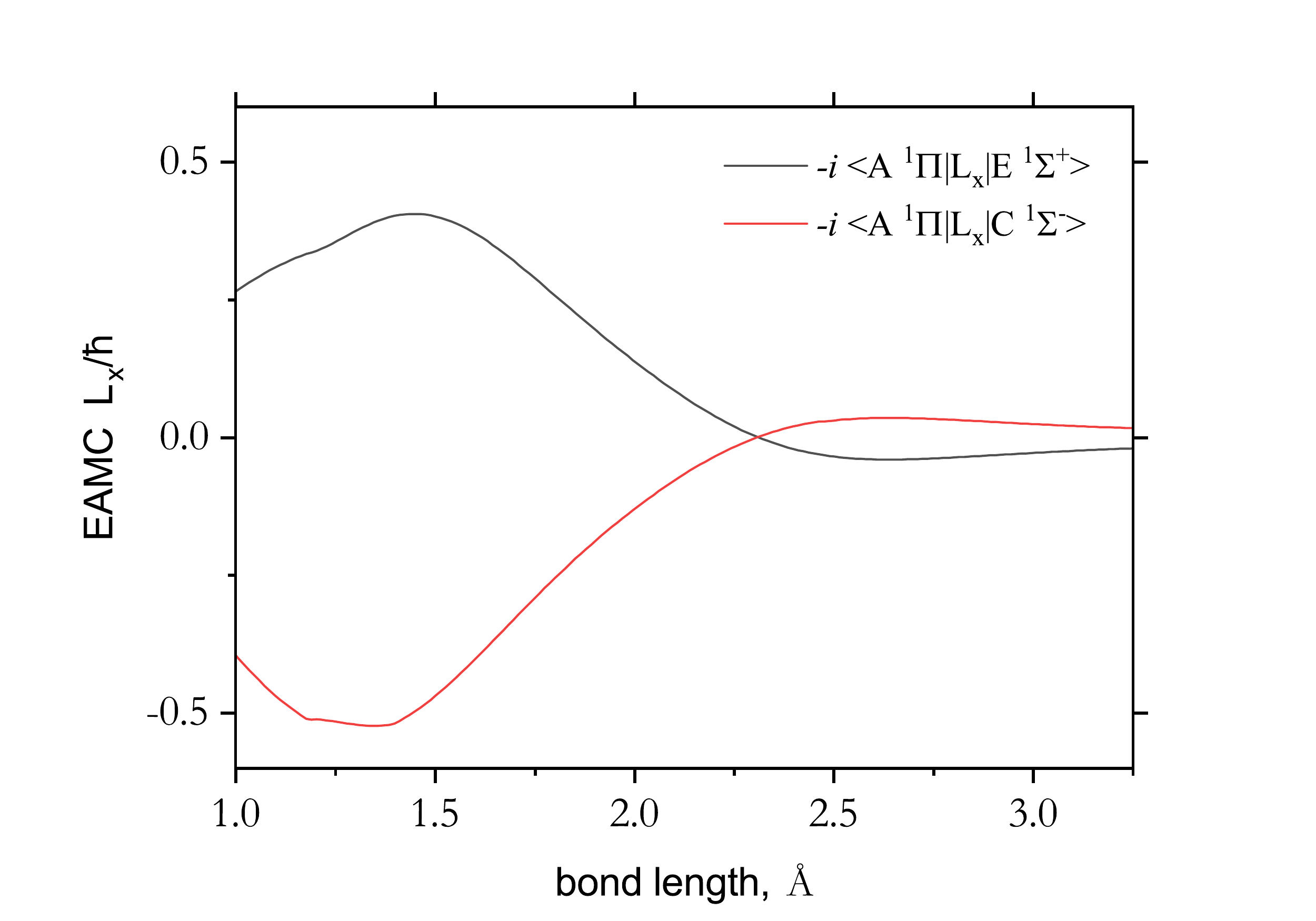}
    \includegraphics[width=0.44\textwidth]{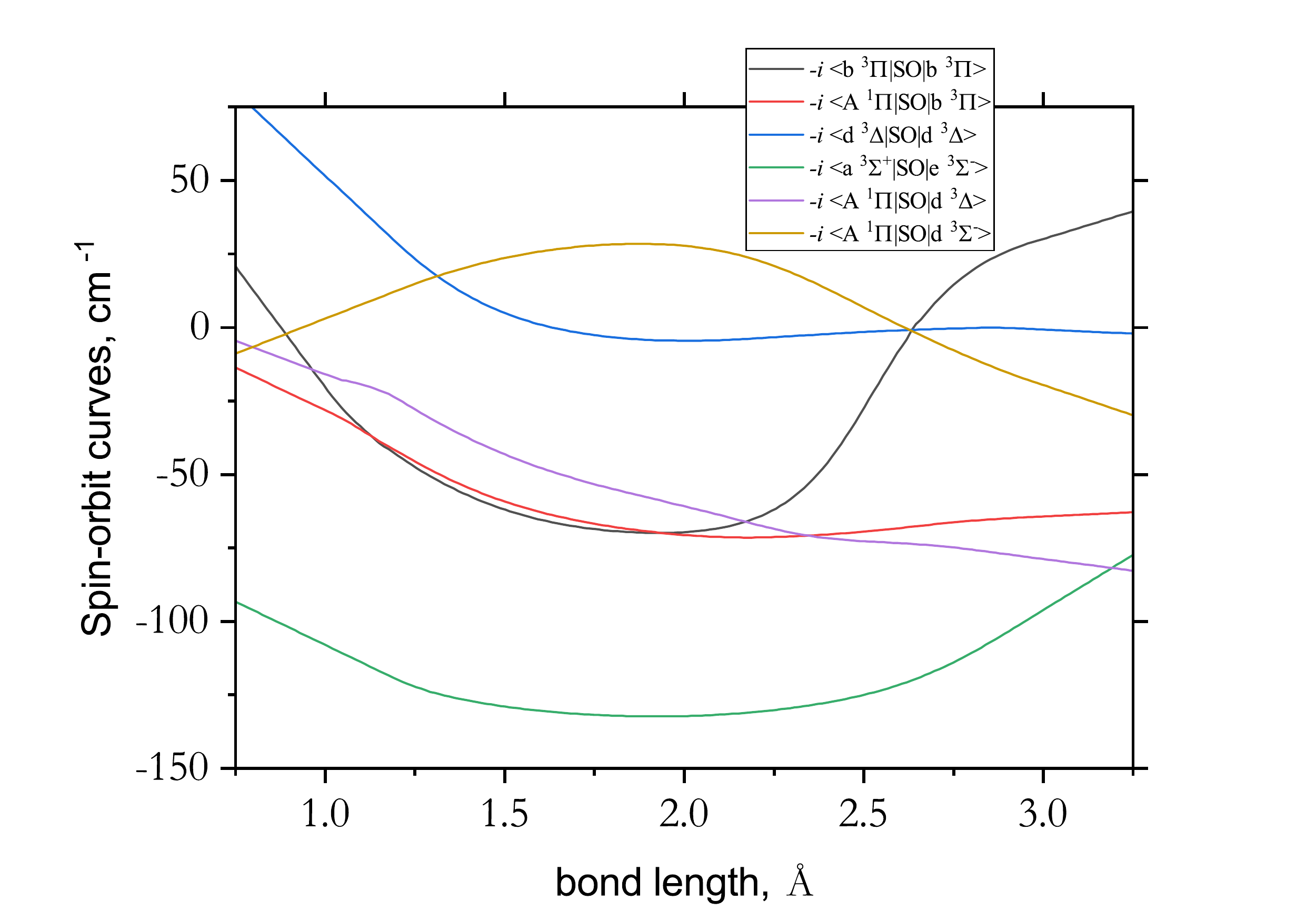}
    \caption{Refined electronic angular momentum and spin-orbit curves for SiO. }
    \label{fig:EAMC:SO}
\end{figure}

The initial singlet  \A, \C, \D, \E\  PECs of SiO were taken from \citet{16Bauschlicher.SiO}, see Fig.~\ref{fig:PECs}.
%\red{JT: I am completely confused about what is actually used from our calculations. One reading would be nothing and they are irrelevant or is the triplet curves or what?. Can some please specify?} %\blue{Yes, the triplets are from our calculations}
The initial  PECs for the triplet states (Fig.~\ref{fig:PECs}) and initial SOCs and EAMCs  shown in Fig.~\ref{fig:EAMC:SO}  were computed  \ai\ in this work using the MRCI/aug-cc-pV5Z  level of theory. The \ai\ curves were then refined by fitting to MARVEL energies.  For the \X\ PEC we used the empirical curve from \citet{jt563}.

The \ai\ calculations  in this work were performed using the internally contracted multireference configuration interaction (IC-MRCI-F12c) approach
with the F12-optimised correlation consistent basis set, {QZ-F12}~\citep{08PeAdWe.ai}  in the frozen core approximation. The active space and state-averaging was chosen as in \citet{16Bauschlicher.SiO}: occupied (8,3,3,0), closed (5,1,1,0) with two $A_1$ and one $\Pi$ states.

Calculations employed the diagonal fixed amplitude ansatz 3C(FIX)~\citep{04TenNo.ai} and a Slater geminal exponent value of {$\beta=1.0$~$a_0^{-1}$}~\citep{09HiPeKn.ai}.  MOLPRO2015~\citep{MOLPRO} was used for all electronic structure calculations. A dense grid of {130} Si--O   bond lengths  was used.

In \textsc{Duo} all PECs  were represented using an Extended Morse Oscillator (EMO) function \citep{EMO} as given by
\begin{equation}\label{e:EMO}
V(r)=V_{\rm e}\;\;+\;\;(A_{\rm e} - V_{\rm
e})\left[1\;\;-\;\;\exp\left(-\sum_{k=0}^{N} B_{k}\xi_p^{k}(r-r_{\rm e})
\right)\right]^2,
\end{equation}
where $A_{\rm e} - V_{\rm e} = D_{\rm e}$ is the dissociation energy, $A_{\rm e}$ is the corresponding asymptote, $r_{\rm e}$ is an equilibrium distance of the PEC, and $\xi_p$ is the \v{S}urkus variable \citep{84SuRaBo.method} given by
\begin{equation}
\label{e:surkus:2}
\xi_p= \frac{r^{p}-r^{p}_{\rm e}}{r^{p}+r^{p}_{\rm e }}
\end{equation}
with $V_{\rm e} = 0$ for the \X\ state.
All  states except  \A\ have a common dissociation limit which we fixed to the ground state dissociation energy $D_{\rm e}$ {8.26}~eV given by  \citet{79HeHuxx.book}; this value was used to define the $X$ PEC  in \citep{jt563}.  A mass spectrometric dissociation energy   by   \citet{69HiMuxx.SiO} gives $D_{0}^0 = $7.93~eV while \citet{79LaArxx.SiO} obtained 8.1 eV computed using  an  \ai\ configuration interaction method.

The spin-orbit curves as the EAMC of $A$--$E$ were morphed using the expansion
\begin{equation}
\label{e:bob}
F(r)=\sum^{N}_{k=0}B_{k}\, z^{k} (1-\xi_p) + \xi_p\, B_{\infty},
\end{equation}
where $z$ is the
damped-coordinate polynomial given by (for SOCs, EAMCs $A$--$B$, SRCs):
\begin{equation}\label{e:damp}
z = (r-r_{\rm ref})\, e^{-\beta_2 (r-r_{\rm ref})^2-\beta_4 (r - r_{\rm ref})^4},
\end{equation}
see also \citet{jt703} and \citet{jt711}.  Here $r_{\rm ref}$ is a
reference position chosen to be close to $r_{\rm e}$ of \X\ and $\beta_2$ and
$\beta_4$ are damping factors.

%There are several thermochemical D0 0 values available for SiO. The calculated De is found to be 7.87 eV, which compares well with the experimental values of 7.93 ( 0.13,23 8.18 ( 0.3) (Gaydon, A. G. Dissociation Energies and Spectra of Diatomic Molecules; Chapman and Hall: London, 1968)  and 8.36 eV. ( Brewer, L.; Rosenblatt, G. M. AdVances in High-Temperature Chemistry; Academic: New York, 1969; Vol. 2) The smallest discrepancy of about 0.3 eV may be considered to be within the limit of accuracy of the MRDCI methodology used here. However, the ground state De of 8.1 eV computed from the CI calculations by Langhoff and Arnold 79LaArxx.SiO has shown a better agreement with the ex

\subsection{Dipole moment curves}

\begin{figure}
    \centering
    \includegraphics[width=0.80\textwidth]{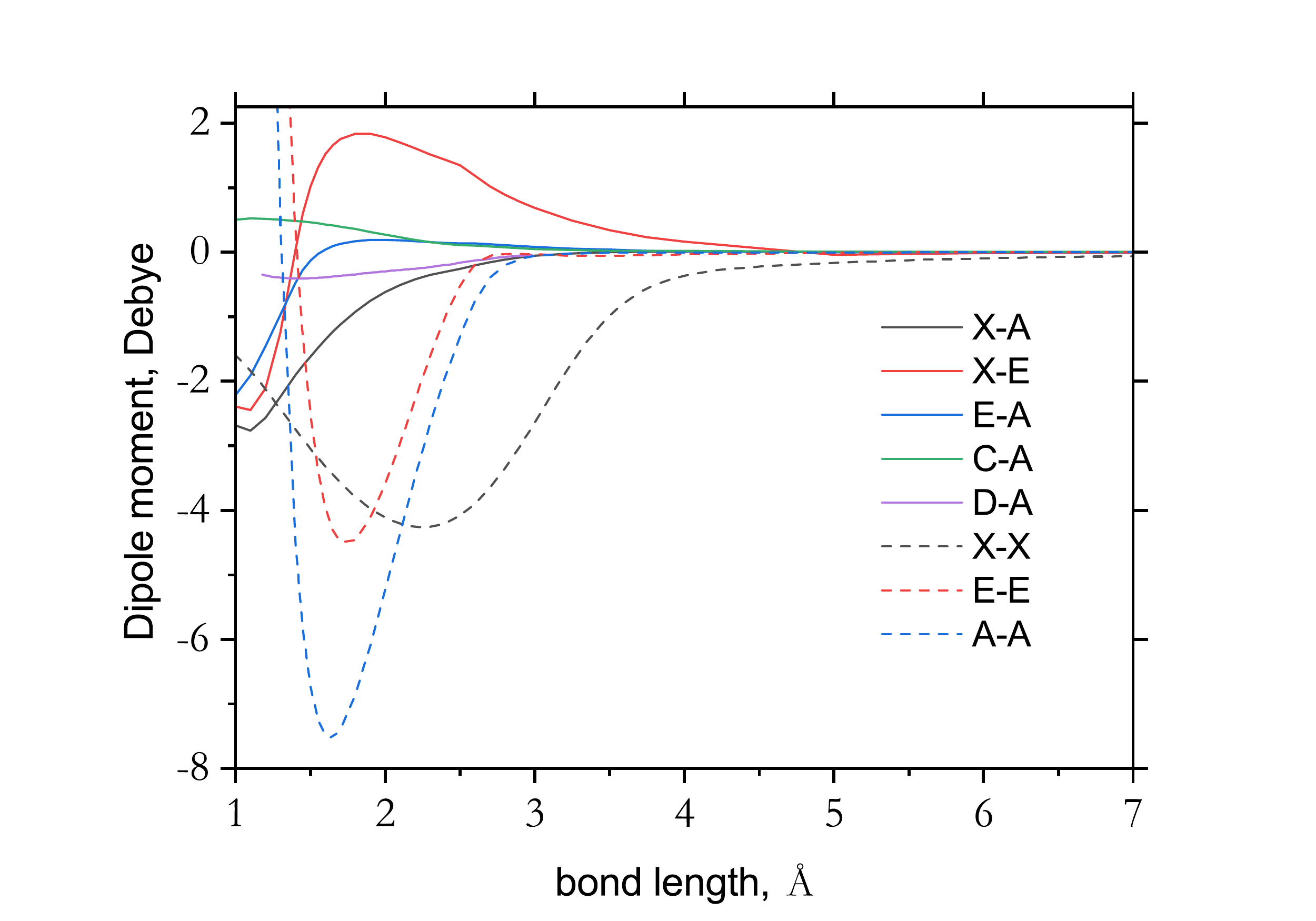}
    \caption{ \textit{Ab initio} (transition) dipole moment curves used in this work are from \citet{16Bauschlicher.SiO} except $X$--$X$ which is taken from \citep{jt563} and the  \D-\A\  TDMC which is computed in this work.}
    \label{fig:DMCs}
\end{figure}

The dipole moment curve of \X\ was taken from \citet{jt563}.
The (transition) DMCs \A--\X, \E--\X\ (most important for this study), \A--\A, \E--\E, \E--\A, \C--\A\  were taken from \citet{16Bauschlicher.SiO}, and the  \D-\A\  TDMC has been computed in this work. The phases of these non-diagonal TDMCs were selected to be consistent with the phases of the \ai\ curves produced in our MOLPRO calculations.
The \ai\ (transitional) dipole moment curves used in this work are shown in Fig.~\ref{fig:DMCs}.

The \ai\ DMC of \X\ was modelled using the expansion
\begin{equation}
\label{e:DMC}
\mu(r)= (1-\xi_p) \sum^{N}_{k=0}C_{k}\, z^{k}  + \xi_p\, C_{\infty},
\end{equation}
where $z$ is  taken as the damped-coordinate $z$ from Eq.~\eqref{e:damp}, see also \citet{jt703} and \citet{jt711}. This was done to  reduce the numerical errors commonly appearing in simulations of spectra of high overtones, see \citet{16MeMeSt}.
Here   $C_k$ and  $C_{\rm \infty}$ are adjustable parameters. The expansion centre $r_{\rm ref}$ is typically chosen to be close to the equilibrium value of the ground electronic state. The $X$ state dipole moment curve from the EBJT model  was transformed to the analytical representation of Eq.~\eqref{e:damp}.

The vibrational transitional dipole moment for the ground ($X$) vibrational state $v=0$ is {-3.0803~D}, which compares well with the experimentally determined values of 3.0982 ($D_0$) and 3.088~D ($D_e$)  using the molecular beam electric resonance spectra  \citep{70RaMuKl.SiO}. In Table~\ref{t:SiO:vib-dm} we compare diagonal vibrational transition dipole moments with the experimental values by  \citep{70RaMuKl.SiO} for a few  $v'=v''$ vibrational  bands, showing excellent agreement.

\begin{table*}
\caption{Comparison of experimental  (molecular beam electric resonance spectra  \citep{70RaMuKl.SiO}) and theoretical (this work) diagonal vibrational transition dipole moments (Debye) of SiO in its ground $X$ electronic state for $v'=v'=0-4$.  }
\label{t:SiO:vib-dm}
    \begin{tabular}{rrr}
        \hline \hline
        Band & Exp. (D) & This work (D) \\
        \hline
0--0 &  3.098   &       3.080   \\
1--1 &  3.118   &       3.100   \\
2--2 &  3.118   &       3.120   \\
3--3 &  3.137   &       3.139   \\
4--4 &  3.157   &       3.159   \\
\hline \hline
\end{tabular}
\end{table*}

% The \X-state PEC is taken from the previous ExoMol model unchanged.
% Refined PECs, EAMCs and SOCS are shown in Figs.~\ref{fig:PECs}--\ref{fig:EAMC:SO}.

All expansion parameters or curves defining our spectroscopic model  are given as supplementary material to the paper as a \textsc{Duo} input file.

\section{Deperturbation}
\label{sec:duo}

In \textsc{Duo} calculations a grid of 501 Sinc DVR points ranging from 0.2 to 7~\AA\ was used. The vibrational basis set consisted of 40, 40, 50, 30, 30, 30, 30, 30, 30  vibrational basis functions for the \X, \A, \E, \C, \D, \astate, \bstate, \cstate\ and \dstate\ states, respectively, generated by solving the three independent $J=0$ Schr\"{o}dinger equations using the Sinc DVR method \citep{jt609}.

\begin{figure}
    \centering
    \includegraphics{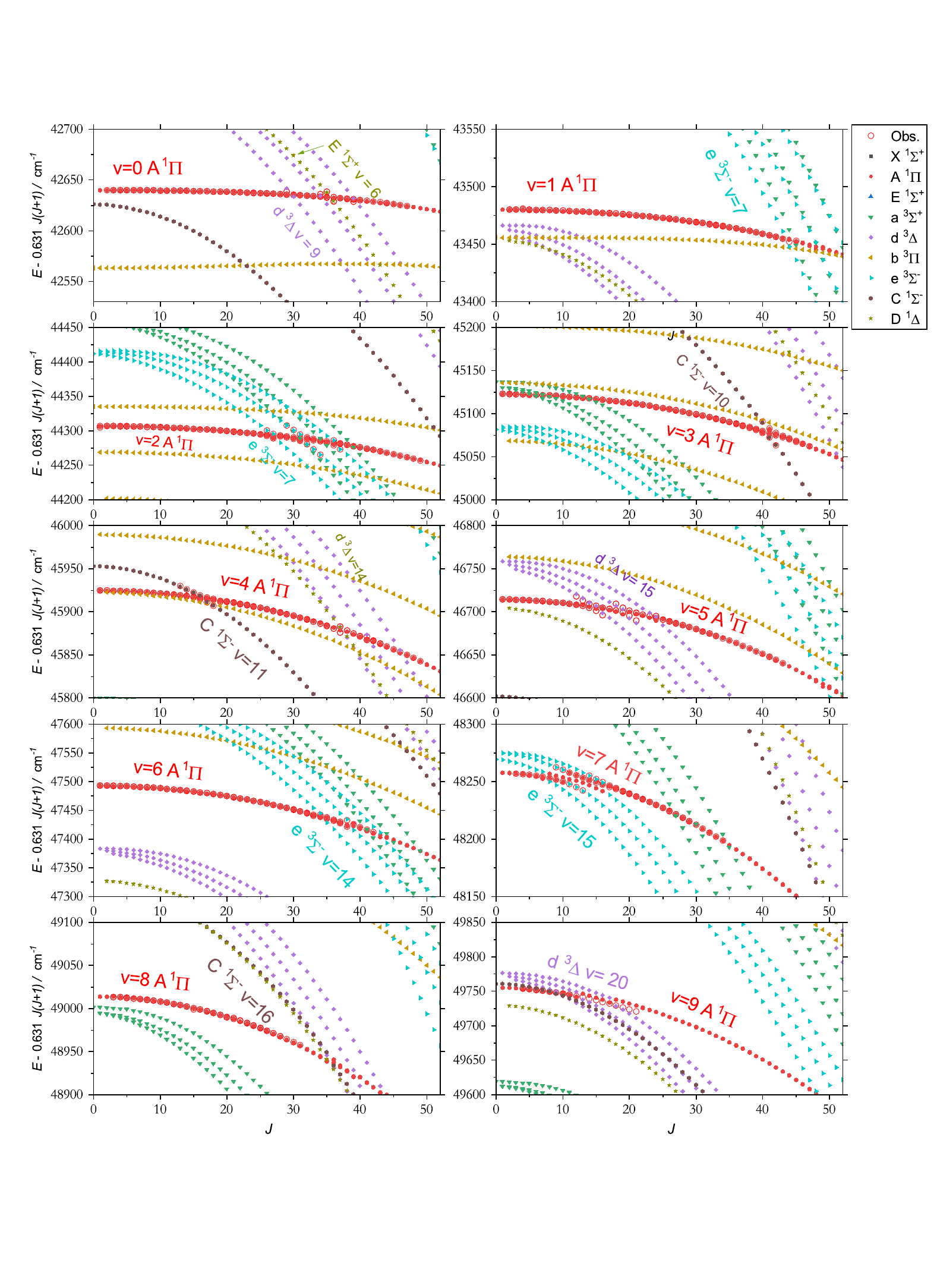}
    \caption{Perturbations of the \A\ state term values by  `dark' sates. Reduced energy term values $E- 0.631 J(J+1)$  are shown as a function of $J$. The red open   circles represent experimental levels from  the  $A$ system. The filled symbols represent the calculated (\textsc{Duo}) levels with the red circles indicating the $A$ states. }
    \label{fig:A:energies}
\end{figure}

Figure~\ref{fig:A:energies} illustrates the main perturbations of the \A\ system $(v=0,\ldots,9)$ caused by the nearby lying electronic states \C, \D, \astate, \bstate, \cstate\ and \dstate. This figures shows all empirical  and theoretical (\textsc{Duo}) term values as a function of $J$, reduced as
$$
\tilde{E}_{i}^{\rm red} = \tilde{E}_{i} - 0.631 J (J+1)
$$
to remove the major $J(J+1)$ contribution.

%The experiment and analysis by \citet{76FiLaRe.SiO} is  supplemented by  transitions from the $0-1$, $0-2$ and $0-3$ band of the   $A$--$X$ system enhanced through perturbations with $^1\Delta$ and $^3\Sigma^-$ as reported by  \citet{76BrReCo.SiO} \red{check}.

The empirical term values representing the $A$--$X$ system were obtained by combing the experimental transition frequencies by \citet{76FiLaRe.SiO,76BrReCo.SiO} and the accurate lower state $X$ of SiO as taken from the ExoMol line list EBJT for SiO. All perturbing states are dark, i.e. electric dipole   transitions to and from the \X\ state are forbidden, and appear in the experiment only because of the  interaction with the `bright' \A\ state, allowed for \X, via the so-called intensity borrowing.  These interactions are very local to the crossing, giving rise to more transitions than  would otherwise be expected for the $A$--$X$ system.  The $J$ dependencies of the energies around each crossing break the rovibronic sequences  of the $A$ term values. The shape of the crossing is very specific to the character of the crossing state, not only to its multiplicity, but also to the curvature exhibiting the differences between the corresponding rotational constants: the larger the difference of $B$ from 0.631~\cm, the steeper the $J$ curve. These shapes represent signatures of the state in question and are used to reconstruct their relative positions on the energy diagrams as in Fig.~\ref{fig:A:energies}.

\citet{76FiLaRe.SiO} reported a detailed analysis of all perturbations that appeared in the $A$--$X$ spectra they observed, including assignments of the crossing rovibronic energies and estimates of the corresponding spectroscopic constants. This information was extremely valuable for our analysis. We first used their spectroscopic constants to reconstruct rovibronic energies of the perturbing state \C, \D, \astate, \bstate, \cstate\ and \dstate\ around crossing points with the program \textsc{PGOPHER} \citep{PGOPHER} and then to initially refine the corresponding \ai\ curves by fitting to these  empirical energies using \textsc{Duo}. This allowed us to correlate  the \textsc{Duo} energy levels to the experimental (MARVEL) values and properly assign some of the duplicate transitions from \citet{76BrReCo.SiO} and \citet{76FiLaRe.SiO} to forbidden vibronic systems $C$--$X$, $D$--$X$, $c$--$X$ and $d$--$X$, with the help of the detailed description of interactions  provided by \citet{76FiLaRe.SiO}. This was important because the \ai\ spectroscopic model (PECs and SOCs) is not accurate enough for predicting the exact  vibronic state, especially when  the corresponding vibrational quantum numbers are very high ($v=10$--20).  The re-assignments of 112 transitions from \citet{76BrReCo.SiO} and \citet{76FiLaRe.SiO}  are listed in Tables \ref{t:76BrReCo:new} and \ref{t:76FiLaRe:new}.

%\red{JT: These new (or proper) assignments are important. If there are too many, the abstract says 112, to give as a table in main text then we should certainly provide a seperate table of them in the supplementary material.}

%The experiment and analysis by \citet{76FiLaRe.SiO} is  supplemented by  transitions from the $0-1$, $0-2$ and $0-3$ band of the   $A$--$X$ system enhanced through perturbations with $^1\Delta$ and $^3\Sigma^-$ as reported by  \citet{76BrReCo.SiO} \red{check}.

There are no experimental line positions for the \astate\ state. \citet{75HaHaHa.SiO} estimated  $T_0$ as 33\,409 \cm, which we used to adjust the corresponding $T_{\rm e}$ constant of this state.

\begin{table*}
\caption{ Assignment of forbidden  transitions from \citet{76BrReCo.SiO}.}
\label{t:76BrReCo:new}
\begin{tabular}{rlrrcrrrcrlrrcrrr}
        \hline \hline
         $\tilde\nu$ (\cm)  &  State'   &  $J'$       &  $v'$       &    e/f' &  $\Omega'$ &  $J"$   & $ v''$     & & $\tilde\nu$ (\cm)  &  State'   &  $J'$       &  $v'$       &    e/f' &  $\Omega'$ &  $J"$   & $ v''$      \\
         \hline
    41349.36 &  $D{}^1 \Delta$   &   35 &    6 &   e  &    2 &   34 &    1  &    &  40160.70  &   $d{}^3 \Delta$ &    29 &     9 &   e   &     2 &    28 &     2   \\
    41247.78 &  $D{}^1 \Delta$   &   35 &    6 &   e  &    2 &   36 &    1  &    &  40076.77  &   $d{}^3 \Delta$ &    29 &     9 &   e   &     2 &    30 &     2   \\
    41299.36 &  $D{}^1 \Delta$   &   35 &    6 &   f  &    2 &   35 &    1  &    &  40119.35  &   $d{}^3 \Delta$ &    29 &     9 &   f   &     2 &    29 &     2   \\
    40137.54 &  $D{}^1 \Delta$   &   35 &    6 &   e  &    2 &   34 &    2  &    &  38958.90  &   $d{}^3 \Delta$ &    29 &     9 &   e   &     2 &    28 &     3   \\
    40036.45 &  $D{}^1 \Delta$   &   35 &    6 &   e  &    2 &   36 &    2  &    &  38875.30  &   $d{}^3 \Delta$ &    29 &     9 &   e   &     2 &    30 &     3   \\
    40087.69 &  $D{}^1 \Delta$   &   35 &    6 &   f  &    2 &   35 &    2  &    &  38917.80  &   $d{}^3 \Delta$ &    29 &     9 &   f   &     2 &    29 &     3   \\
    38937.63 &  $D{}^1 \Delta$   &   35 &    6 &   e  &    2 &   34 &    3  &    &  41351.85  &   $d{}^3 \Delta$ &    34 &     9 &   e   &     1 &    33 &     1   \\
    38837.16 &  $D{}^1 \Delta$   &   35 &    6 &   e  &    2 &   36 &    3  &    &  41252.76  &   $d{}^3 \Delta$ &    34 &     9 &   e   &     1 &    35 &     1   \\
    38888.06 &  $D{}^1 \Delta$   &   35 &    6 &   f  &    2 &   35 &    3  &    &  41302.91  &   $d{}^3 \Delta$ &    34 &     9 &   f   &     1 &    34 &     1   \\
    41335.65 &  $D{}^1 \Delta$   &   36 &    6 &   e  &    2 &   35 &    1  &    &  40139.25  &   $d{}^3 \Delta$ &    34 &     9 &   e   &     1 &    33 &     2   \\
    41230.94 &  $D{}^1 \Delta$   &   36 &    6 &   e  &    2 &   37 &    1  &    &  40040.92  &   $d{}^3 \Delta$ &    34 &     9 &   e   &     1 &    35 &     2   \\
    41283.97 &  $D{}^1 \Delta$   &   36 &    6 &   f  &    2 &   36 &    1  &    &  40090.85  &   $d{}^3 \Delta$ &    34 &     9 &   f   &     1 &    34 &     2   \\
    40123.76 &  $D{}^1 \Delta$   &   36 &    6 &   e  &    2 &   35 &    2  &    &  38938.91  &   $d{}^3 \Delta$ &    34 &     9 &   e   &     1 &    33 &     3   \\
    40020.00 &  $D{}^1 \Delta$   &   36 &    6 &   e  &    2 &   37 &    2  &    &  38841.40  &   $d{}^3 \Delta$ &    34 &     9 &   e   &     1 &    35 &     3   \\
    40072.57 &  $D{}^1 \Delta$   &   36 &    6 &   f  &    2 &   36 &    2  &    &  38890.89  &   $d{}^3 \Delta$ &    34 &     9 &   f   &     1 &    34 &     3   \\
    38924.43 &  $D{}^1 \Delta$   &   36 &    6 &   e  &    2 &   35 &    3  &    &  41209.05  &   $d{}^3 \Delta$ &    39 &     9 &   e   &     0 &    40 &     1   \\
    38821.02 &  $D{}^1 \Delta$   &   36 &    6 &   e  &    2 &   37 &    3  &    &  41266.66  &   $d{}^3 \Delta$ &    39 &     9 &   f   &     0 &    39 &     1   \\
    38873.36 &  $D{}^1 \Delta$   &   36 &    6 &   f  &    2 &   36 &    3  &    &  40111.69  &   $d{}^3 \Delta$ &    39 &     9 &   e   &     0 &    38 &     2   \\
    41374.54 &  $d{}^3 \Delta$   &   29 &    9 &   e  &    2 &   28 &    1  &    &  39999.18  &   $d{}^3 \Delta$ &    39 &     9 &   e   &     0 &    40 &     2   \\
    41289.93 &  $d{}^3 \Delta$   &   29 &    9 &   e  &    2 &   30 &    1  &    &  40056.34  &   $d{}^3 \Delta$ &    39 &     9 &   f   &     0 &    39 &     2   \\
    41332.80 &  $d{}^3 \Delta$   &   29 &    9 &   f  &    2 &   29 &    1  &    &  41322.36  &   $d{}^3 \Delta$ &    39 &     9 &   e   &     0 &    38 &     1   \\
    \hline\hline
\end{tabular}
\end{table*}

\begin{table*}
\caption{ Assignment of forbidden  transitions from \citet{76FiLaRe.SiO}.}
\label{t:76FiLaRe:new}
\begin{tabular}{rlrrcrrrcrlrrcrrr}
        \hline \hline
         $\tilde\nu$ (\cm)  &  State'   &  $J'$       &  $v'$       &    e/f' &  $\Omega'$ &  $J"$   & $ v''$     & & $\tilde\nu$ (\cm)  &  State'   &  $J'$       &  $v'$       &    e/f' &  $\Omega'$ &  $J"$   & $ v''$      \\
         \hline
    45912.81 &  $C{}^1\Sigma^-$  &   13 &   11 &   f  &    0 &   13 &    0  &    &  46677.51  &   $d{}^3 \Delta$ &    21 &    15 &   e   &     1 &    20 &     0   \\
    45906.48 &  $C{}^1\Sigma^-$  &   14 &   11 &   f  &    0 &   14 &    0  &    &  46647.05  &   $d{}^3 \Delta$ &    21 &    15 &   f   &     1 &    21 &     0   \\
    45899.97 &  $C{}^1\Sigma^-$  &   15 &   11 &   f  &    0 &   15 &    0  &    &  46615.12  &   $d{}^3 \Delta$ &    21 &    15 &   e   &     1 &    22 &     0   \\
    45893.63 &  $C{}^1\Sigma^-$  &   16 &   11 &   f  &    0 &   16 &    0  &    &  46673.81  &   $d{}^3 \Delta$ &    24 &    15 &   e   &     3 &    23 &     0   \\
    45882.76 &  $C{}^1\Sigma^-$  &   17 &   11 &   f  &    0 &   17 &    0  &    &  46639.16  &   $d{}^3 \Delta$ &    24 &    15 &   f   &     3 &    24 &     0   \\
    45875.02 &  $C{}^1\Sigma^-$  &   18 &   11 &   f  &    0 &   18 &    0  &    &  46602.92  &   $d{}^3 \Delta$ &    24 &    15 &   e   &     3 &    25 &     0   \\
    44940.32 &  $C{}^1\Sigma^-$  &   40 &   10 &   f  &    0 &   40 &    0  &    &  45753.88  &   $d{}^3 \Delta$ &    37 &    14 &   f   &     2 &    37 &     0   \\
    44914.70 &  $C{}^1\Sigma^-$  &   41 &   10 &   f  &    0 &   41 &    0  &    &  45698.94  &   $d{}^3 \Delta$ &    37 &    14 &   e   &     2 &    38 &     0   \\
    44898.36 &  $C{}^1\Sigma^-$  &   42 &   10 &   f  &    0 &   42 &    0  &    &  42543.94  &   $d{}^3 \Delta$ &    39 &     9 &   e   &     0 &    38 &     0   \\
    42453.07 &  $D{}^1 \Delta$   &   36 &    6 &   e  &    2 &   37 &    0  &    &  42487.91  &   $d{}^3 \Delta$ &    39 &     9 &   f   &     0 &    39 &     0   \\
    42558.45 &  $D{}^1 \Delta$   &   36 &    6 &   e  &    2 &   35 &    0  &    &  45713.65  &   $d{}^3 \Delta$ &    41 &    14 &   f   &     1 &    41 &     0   \\
    42506.71 &  $D{}^1 \Delta$   &   36 &    6 &   f  &    2 &   36 &    0  &    &  48254.23  &   $e{}^3\Sigma^-$ &     9 &    15 &   f   &     1 &     9 &     0   \\
    46720.58 &  $d{}^3 \Delta$   &   12 &   15 &   e  &    3 &   11 &    0  &    &  48249.98  &   $e{}^3\Sigma^-$ &    10 &    15 &   f   &     1 &    10 &     0   \\
    46703.19 &  $d{}^3 \Delta$   &   12 &   15 &   f  &    3 &   12 &    0  &    &  48245.56  &   $e{}^3\Sigma^-$ &    11 &    15 &   f   &     1 &    11 &     0   \\
    46684.37 &  $d{}^3 \Delta$   &   12 &   15 &   e  &    3 &   13 &    0  &    &  48230.81  &   $e{}^3\Sigma^-$ &    12 &    15 &   f   &     1 &    12 &     0   \\
    46715.29 &  $d{}^3 \Delta$   &   13 &   15 &   e  &    2 &   12 &    0  &    &  48225.27  &   $e{}^3\Sigma^-$ &    13 &    15 &   f   &     1 &    13 &     0   \\
    46696.51 &  $d{}^3 \Delta$   &   13 &   15 &   f  &    2 &   13 &    0  &    &  48255.85  &   $e{}^3\Sigma^-$ &    14 &    15 &   e   &     1 &    13 &     0   \\
    46676.19 &  $d{}^3 \Delta$   &   13 &   15 &   e  &    2 &   14 &    0  &    &  48213.84  &   $e{}^3\Sigma^-$ &    14 &    15 &   e   &     1 &    15 &     0   \\
    46710.10 &  $d{}^3 \Delta$   &   14 &   15 &   e  &    2 &   13 &    0  &    &  48251.55  &   $e{}^3\Sigma^-$ &    15 &     7 &   e   &     0 &    14 &     0   \\
    46689.90 &  $d{}^3 \Delta$   &   14 &   15 &   f  &    2 &   14 &    0  &    &  48206.65  &   $e{}^3\Sigma^-$ &    15 &    15 &   e   &     0 &    16 &     0   \\
    46668.11 &  $d{}^3 \Delta$   &   14 &   15 &   e  &    2 &   15 &    0  &    &  48247.24  &   $e{}^3\Sigma^-$ &    16 &    15 &   e   &     1 &    15 &     0   \\
    46700.44 &  $d{}^3 \Delta$   &   15 &   15 &   e  &    3 &   14 &    0  &    &  48199.56  &   $e{}^3\Sigma^-$ &    16 &    15 &   e   &     1 &    17 &     0   \\
    46678.73 &  $d{}^3 \Delta$   &   15 &   15 &   f  &    3 &   15 &    0  &    &  44272.94  &   $e{}^3\Sigma^-$ &    26 &     9 &   e   &     0 &    25 &     0   \\
    46655.61 &  $d{}^3 \Delta$   &   15 &   15 &   e  &    3 &   16 &    0  &    &  44264.79  &   $e{}^3\Sigma^-$ &    27 &     9 &   e   &     0 &    26 &     0   \\
    46670.75 &  $d{}^3 \Delta$   &   16 &   15 &   f  &    2 &   16 &    0  &    &  44227.74  &   $e{}^3\Sigma^-$ &    29 &     9 &   f   &     1 &    29 &     0   \\
    46646.05 &  $d{}^3 \Delta$   &   16 &   15 &   e  &    2 &   17 &    0  &    &  44215.43  &   $e{}^3\Sigma^-$ &    30 &     9 &   f   &     1 &    30 &     0   \\
    46703.19 &  $d{}^3 \Delta$   &   18 &   15 &   e  &    3 &   17 &    0  &    &  44203.50  &   $e{}^3\Sigma^-$ &    31 &     9 &   f   &     1 &    31 &     0   \\
    46677.20 &  $d{}^3 \Delta$   &   18 &   15 &   f  &    3 &   18 &    0  &    &  44183.19  &   $e{}^3\Sigma^-$ &    32 &     9 &   f   &     1 &    32 &     0   \\
    46649.70 &  $d{}^3 \Delta$   &   18 &   15 &   e  &    3 &   19 &    0  &    &  44170.00  &   $e{}^3\Sigma^-$ &    33 &     9 &   f   &     1 &    33 &     0   \\
    46696.69 &  $d{}^3 \Delta$   &   19 &   15 &   e  &    3 &   18 &    0  &    &  44155.74  &   $e{}^3\Sigma^-$ &    34 &     9 &   f   &     1 &    34 &     0   \\
    46669.20 &  $d{}^3 \Delta$   &   19 &   15 &   f  &    3 &   19 &    0  &    &  44215.43  &   $e{}^3\Sigma^-$ &    36 &     9 &   e   &     1 &    35 &     0   \\
    46640.24 &  $d{}^3 \Delta$   &   19 &   15 &   e  &    3 &   20 &    0  &    &  44110.19  &   $e{}^3\Sigma^-$ &    36 &     9 &   e   &     1 &    37 &     0   \\
    46685.22 &  $d{}^3 \Delta$   &   20 &   15 &   e  &    1 &   19 &    0  &    &  44197.39  &   $e{}^3\Sigma^-$ &    37 &     9 &   e   &     1 &    36 &     0   \\
    46656.19 &  $d{}^3 \Delta$   &   20 &   15 &   f  &    1 &   20 &    0  &    &  44089.08  &   $e{}^3\Sigma^-$ &    37 &     9 &   e   &     1 &    38 &     0   \\
    46625.89 &  $d{}^3 \Delta$   &   20 &   15 &   e  &    1 &   21 &    0  &    &  47296.58  &   $e{}^3\Sigma^-$ &    38 &    14 &   f   &     1 &    38 &     0   \\
    \hline
    \hline
\end{tabular}
\end{table*}

\begin{figure}
    \centering
    \includegraphics[width=0.8\textwidth]{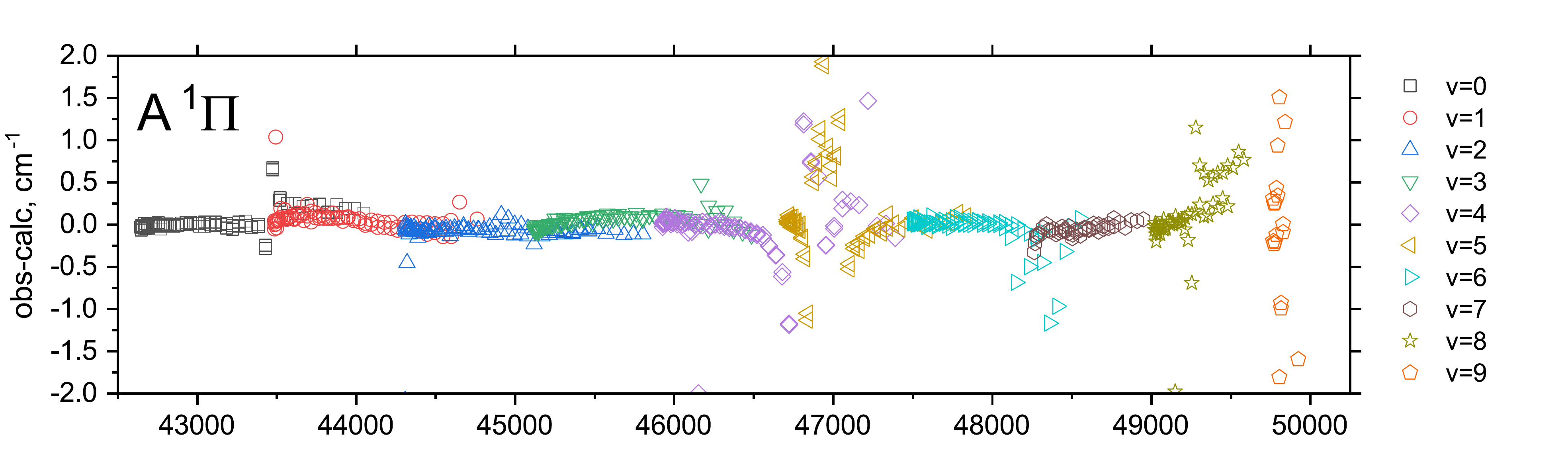}
    \includegraphics[width=0.8\textwidth]{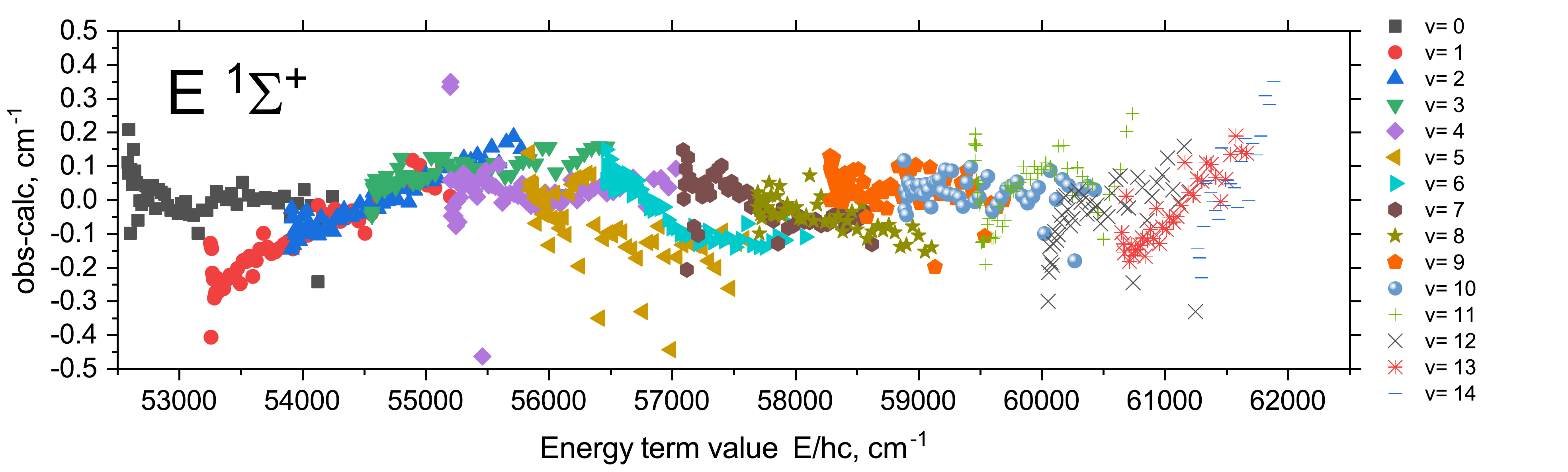}
    \caption{Residuals (Obs.-Calc.) between the experimentally determined energies of SiO (\A\ and \E)  from our MARVEL analysis and the \textsc{Duo} energies corresponding to our refined spectroscopic model.}
    \label{fig:residueAE}
\end{figure}

Our coupled model significantly improves the description of the perturbations. The residual errors between the observed and calculated values (obs$-$calc) are illustrated in Fig.~\ref{fig:residueAE}.  In fitting, we used the EMO form in Eq.~\eqref{e:EMO} to represent the PECs  and the damped form in Eq.~\eqref{e:DMC} to represent the morphing of SOCs of the couplings with the perturbing states. Owing to the scarcity of information on  dark states, only a minimal number of the expansion parameters $B_k$ could be obtained.
The assignment of the dark states around state crossings is illustrated in Fig.~\ref{fig:A:energies}, where the experimental  energy values (open red circles) are overlaid with the theoretical ones. Most of the  experimental points nicely follow the $A$ state sequences, except around the crossing points where these sequences break and follow the corresponding patterns of the dark states, forming  distinct  patterns.  It should be noted that the rovibronic  assignment of interacting levels  exactly at their crossing is always ambiguous owing to the mixed nature of the resonating states.

%\begin{figure}
%    \centering
%    \includegraphics{comparePEC.pdf}
%    \caption{The \textit{ab initio} and
%        refined PECs of $\mathrm{A}\,^1\Pi$ and $\mathrm{E}\,^1\Sigma^+$ states. \red{is it possible to remove the gridlines, to look more like Fig \ref{f:XAE} and \ref{f:temp} ?}}
%    \label{fig:comparePEC}
%\end{figure}

Figure~\ref{fig:residueAE} shows the residuals representing our fit. Some of the $A$ state energies still show small perturbations due to interactions with dark electronic states, which are not fully accounted for in our model.

%\begin{figure}
%    \centering
%    \includegraphics{residueAE1.pdf}
%    \caption{Residuals before inclusion of the 6 %perturbing states: \red{In case we want to show this!}}
%    \label{fig:residueAE1}
%\end{figure}

\section{Line list and simulations of spectra of SiO}
\label{s:linelist}

The \name\ line list  was produced with \textsc{Duo} using the refined spectroscopic model of SiO (provided as supplementary material).   It contains {91\,395\,763} transitions and {174\,250} states for  \X\, \A, \E,  \C, \D, \astate, \bstate, \cstate\ and \dstate,  covering wavenumbers up to {72\,000} \cm\ ($\lambda < 140$ nm) and $J = 0 \ldots  250$.  The line list is provided in State and Transition files, as is customary for the ExoMol format \citep{jt810}.  Extracts from the States and Trans files are shown  in Tables \ref{t:states} and \ref{t:trans}, respectively; the full files are available from \url{www.exomol.com} and CDS. The lines list contains state uncertainties, Land\'{e}-$g$ factors \citep{jt656}, lifetimes \citep{jt624} and a partition function which is very similar to that of EBJT due to the negligible contribution from the energy levels belonging to the excited  electronic states.

The calculated energies were replaced with the MARVEL values (MARVELised), where available. We have used the labels `Ca' and `Ma' in the penultimate column of the States file to indicate if the value is calculated (\textsc{Duo}) or MARVELised, respectively.

The uncertainty values in the States file correspond to two cases: the MARVEL uncertainties are used for MARVELised energies, while for the calculated values the following approximate expression is used:
\begin{equation}
\label{e:unc}
{\rm unc} = a v + b J(J+1),
\end{equation}
where $a$ and $b$ are electronic state dependent constant, defined in Table~\ref{t:unc:a:b}.% \red{Alec: Any explanation for this expression and if accurate?}

\begin{table*}
\centering
\caption{Extract from the states file of the \name\ line list for  \sio.   }
%\tt
\label{t:states}
{\tt  \begin{tabular}{rrrrrrrcclrrrrcr} \hline \hline
$i$ & Energy (\cm) & $g_i$ & $J$ & unc &  $\tau$ & $g$& \multicolumn{2}{c}{Parity} 	& State	& $v$	&${\Lambda}$ &	${\Sigma}$ & $\Omega$ & Label & Calc. \\ \hline
    1666 &  44144.709385 &       7 &      3 &   8.920000   &  9.0168E-03  &    1.1217    &  +   &   f   &     d3Delta    &       11 &    2 &   1   &     3 &  Ca   &     44144.709385 \\
    1667 &  44149.467240 &       7 &      3 &   8.920000   &  2.6619E-04  &    0.4728    &  +   &   f   &     d3Delta    &       11 &    2 &   0   &     2 &  Ca   &     44149.467240 \\
    1668 &  44154.189842 &       7 &      3 &   8.920000   &  1.0477E-04  &    -0.2607   &  +   &   f   &     d3Delta    &       11 &    2 &   -1  &     1 &  Ca   &     44154.189842 \\
    1669 &  44209.302522 &       7 &      3 &   8.920000   &  6.3159E-01  &    0.5301    &  +   &   f   &      b3Pi      &       11 &    1 &   1   &     2 &  Ca   &     44209.302522 \\
    1670 &  44276.546978 &       7 &      3 &   8.920000   &  1.0252E-03  &    0.0905    &  +   &   f   &      b3Pi      &       11 &    1 &   0   &     1 &  Ca   &     44276.546978 \\
    1671 &  44314.520663 &       7 &      3 &   0.050000   &  2.8958E-08  &    0.0835    &  +   &   f   &      A1Pi      &        2 &    1 &   0   &     1 &  Ma   &     44314.597828 \\
    1672 &  44342.823666 &       7 &      3 &   8.920000   &  4.8501E-01  &    -0.0368   &  +   &   f   &      b3Pi      &       11 &    1 &   -1  &     0 &  Ca   &     44342.823666 \\
    1673 &  44421.259298 &       7 &      3 &   7.320000   &  1.7643E-05  &    0.1667    &  +   &   f   &    e3Sigma-    &        9 &    0 &   1   &     1 &  Ca   &     44421.259298 \\
    1674 &  44463.770855 &       7 &      3 &   12.120000  &  1.8929E+00  &    0.5833    &  +   &   f   &    a3Sigma+    &       15 &    0 &   1   &     1 &  Ca   &     44463.770855 \\
    1675 &  44472.524815 &       7 &      3 &   12.120000  &  8.5180E+00  &    -0.4164   &  +   &   f   &    a3Sigma+    &       15 &    0 &   0   &     0 &  Ca   &     44472.524815 \\
    1676 &  44644.518546 &       7 &      3 &   7.320000   &  1.8475E-02  &    0.0000    &  +   &   f   &    C1Sigma-    &        9 &    0 &   0   &     0 &  Ca   &     44644.518546 \\
    1677 &  44787.014146 &       7 &      3 &   7.320000   &  3.0209E-01  &    0.3333    &  +   &   f   &     D1Delta    &        9 &    2 &   0   &     2 &  Ca   &     44787.014146 \\
    1678 &  44814.070066 &       7 &      3 &   9.720000   &  1.1037E-02  &    1.1218    &  +   &   f   &     d3Delta    &       12 &    2 &   1   &     3 &  Ca   &     44814.070066 \\
    1679 &  44818.761972 &       7 &      3 &   9.720000   &  3.3377E-04  &    0.4729    &  +   &   f   &     d3Delta    &       12 &    2 &   0   &     2 &  Ca   &     44818.761972 \\
    1680 &  44823.424247 &       7 &      3 &   9.720000   &  1.3523E-04  &    -0.2609   &  +   &   f   &     d3Delta    &       12 &    2 &   -1  &     1 &  Ca   &     44823.424247 \\
    1681 &  45075.695373 &       7 &      3 &   9.720000   &  6.4266E-01  &    0.5298    &  +   &   f   &      b3Pi      &       12 &    1 &   1   &     2 &  Ca   &     45075.695373 \\
\hline
\hline
%%%%
%%%%
\end{tabular}}
\mbox{}\\
{\flushleft
$i$:   State counting number.     \\
$\tilde{E}$: State energy term values in \cm, MARVEL or Calculated (\textsc{Duo}). \\
$g_i$:  Total statistical weight, equal to ${g_{\rm ns}(2J + 1)}$.     \\
$J$: Total angular momentum.\\
unc: Uncertainty, \cm.\\

$\tau$: Lifetime (s$^{-1}$).\\
$g$: Land\'{e} $g$-factors. \\
$+/-$:   Total parity. \\
%$e/f$:   Rotationless parity. \\
State: Electronic state.\\
$v$:   State vibrational quantum number. \\
$\Lambda$:  Projection of the electronic angular momentum. \\
$\Sigma$:   Projection of the electronic spin. \\
$\Omega$:   Projection of the total angular momentum, $\Omega=\Lambda+\Sigma$. \\
Label: `Ma' is for MARVEL and `Ca' is for Calculated. \\
}
\end{table*}
%%%%%%%%%%%%%%%%%%%%%%%%%%%%%%%%%%%%%%%%%%%%%%%%%%%%%%%%%%%%%%%%%%%%%%%%%%%%%%%%%%%%%%%%%%%%%%%%%%%%%%%%%%%%%%%%%%%%%%%%%%%%%%%%%%%%%%%%%%%%

%%%%%%%%%%%%%%%%%%%%%%%%%%%%%%%%%%%%%%%%%%%%%%%%%%%%%%%%%%%%%%%%%%%%%%%%%%%%%%%%%%%%%%%%%%%%%%%%%%%%%%%%%%%%%%%%%%%%%%%%%%%%%%%%%%%%%%%%%%%%
\begin{table}
\centering
\caption{Extract from the transitions file of the \name\ line list for  \sio.}
\tt
\label{t:trans}
\centering
\begin{tabular}{rrrr} \hline\hline
\multicolumn{1}{c}{$f$}	&	\multicolumn{1}{c}{$i$}	& \multicolumn{1}{c}{$A_{fi}$ (s$^{-1}$)}	&\multicolumn{1}{c}{$\tilde{\nu}_{fi}$} \\ \hline
      18  &       506  &     2.4093E+01  &   2097.041341 \\
     508  &        16  &     8.0507E+00  &   2099.618181 \\
     818  &       506  &     9.6724E+00  &   2100.876729 \\
    1928  &       816  &     1.0376E+01  &   2102.115330 \\
    2358  &      1926  &     1.0772E+01  &   2103.333959 \\
    3528  &      2356  &     1.1030E+01  &   2104.532592 \\
    7157  &      8325  &     1.1301E+01  &   2105.654263 \\
    3958  &      3526  &     1.1212E+01  &   2105.711204 \\
    5128  &      3956  &     1.1350E+01  &   2106.869772 \\
    6727  &      7155  &     1.1372E+01  &   2107.160357 \\
    5558  &      5126  &     1.1458E+01  &   2108.008270 \\
    5557  &      6725  &     1.1455E+01  &   2108.646782 \\
    6728  &      5556  &     1.1546E+01  &   2109.126673 \\
 \hline\hline
\end{tabular} \\ \vspace{2mm}
\rm
\noindent
$f$: Upper  state counting number;\\
$i$:  Lower  state counting number; \\
$A_{fi}$:  Einstein-$A$ coefficient in s$^{-1}$; \\
$\tilde{\nu}_{fi}$: transition wavenumber in \cm.\\
\end{table}
%%%%%%%%%%%%%%%%%%%%%%%%%%%%%%%%%%%%%%%%%%%%%%%%%%%%%%%%%%%%%%%%%%%%%%%%%%%%%%%%%%%%%%%%%%%%%%%%%%%%%%%%%%%%%%%%%%%%%%%%%%%%%%%%%%%%%%%%%%%%

\begin{table}
\centering
\caption{$a$ and $b$ constants defining state dependent uncertainties via Eq. \eqref{e:unc}.  }
\label{t:unc:a:b}
 \begin{tabular}{lrr}
  \hline \hline
 State & $a$ & $b$ \\
 \hline
 \X   &  0.002  & 0.0001  \\
 \A   &  0.5  & 0.001  \\
 \E   &  0.5  & 0.001  \\
 All other   &  0.8  & 0.001  \\
 \hline \hline
\end{tabular}
\end{table}

\begin{figure}
    \centering
    \includegraphics[width=0.80\textwidth]{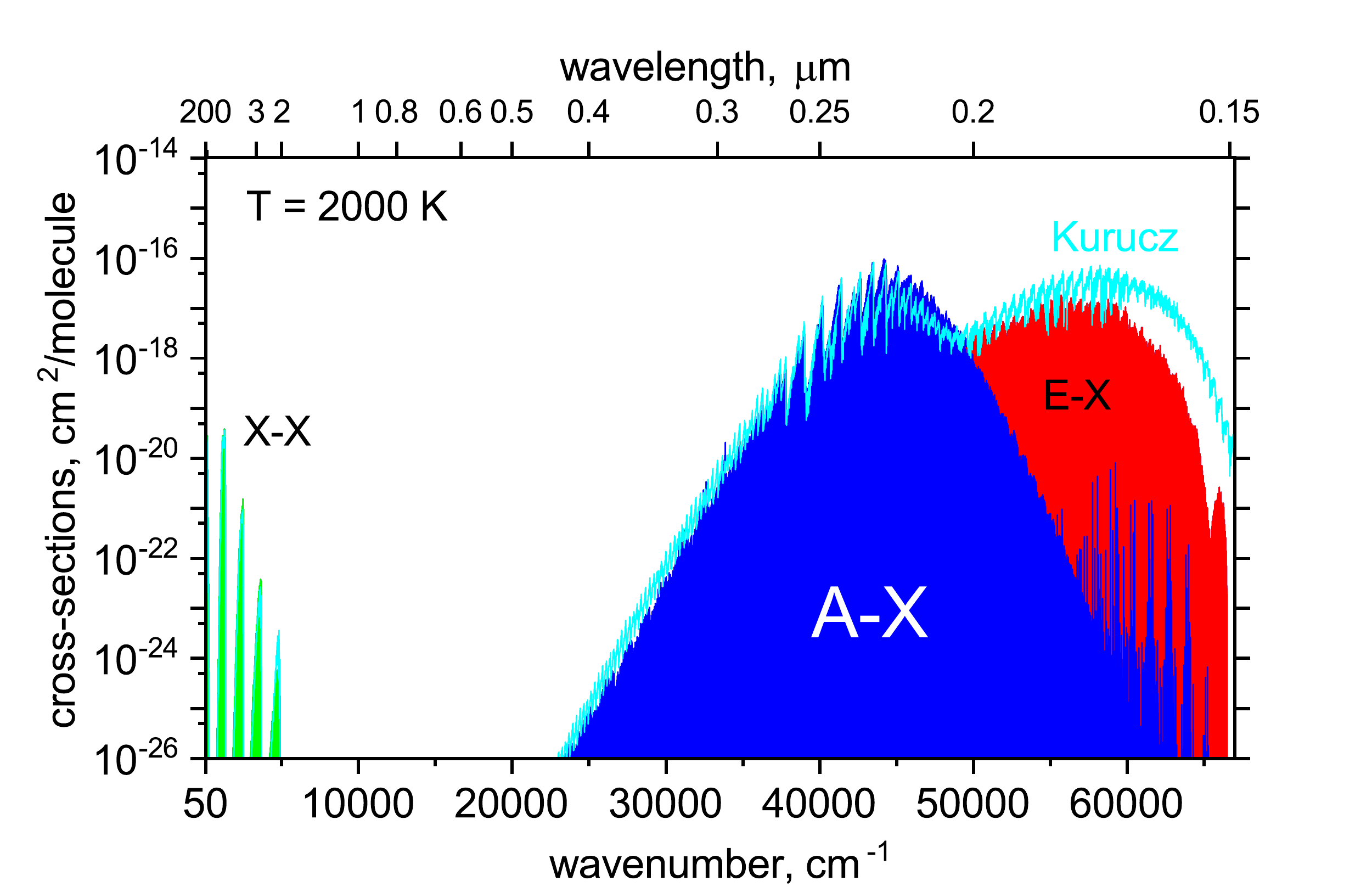}
    \caption{Simulated SiO absorption spectra at 2000 K showing the three main bands of the UV-IR system. A Gaussian profile of half width of half maximum (HWHM) of 1~\cm\ was used. The spectrum based on the line list by \citet{11Kurucz.db} is indicated for comparison. }
    \label{fig:Bands}
\end{figure}

\begin{figure}
    \centering
    \includegraphics[width=0.80\textwidth]{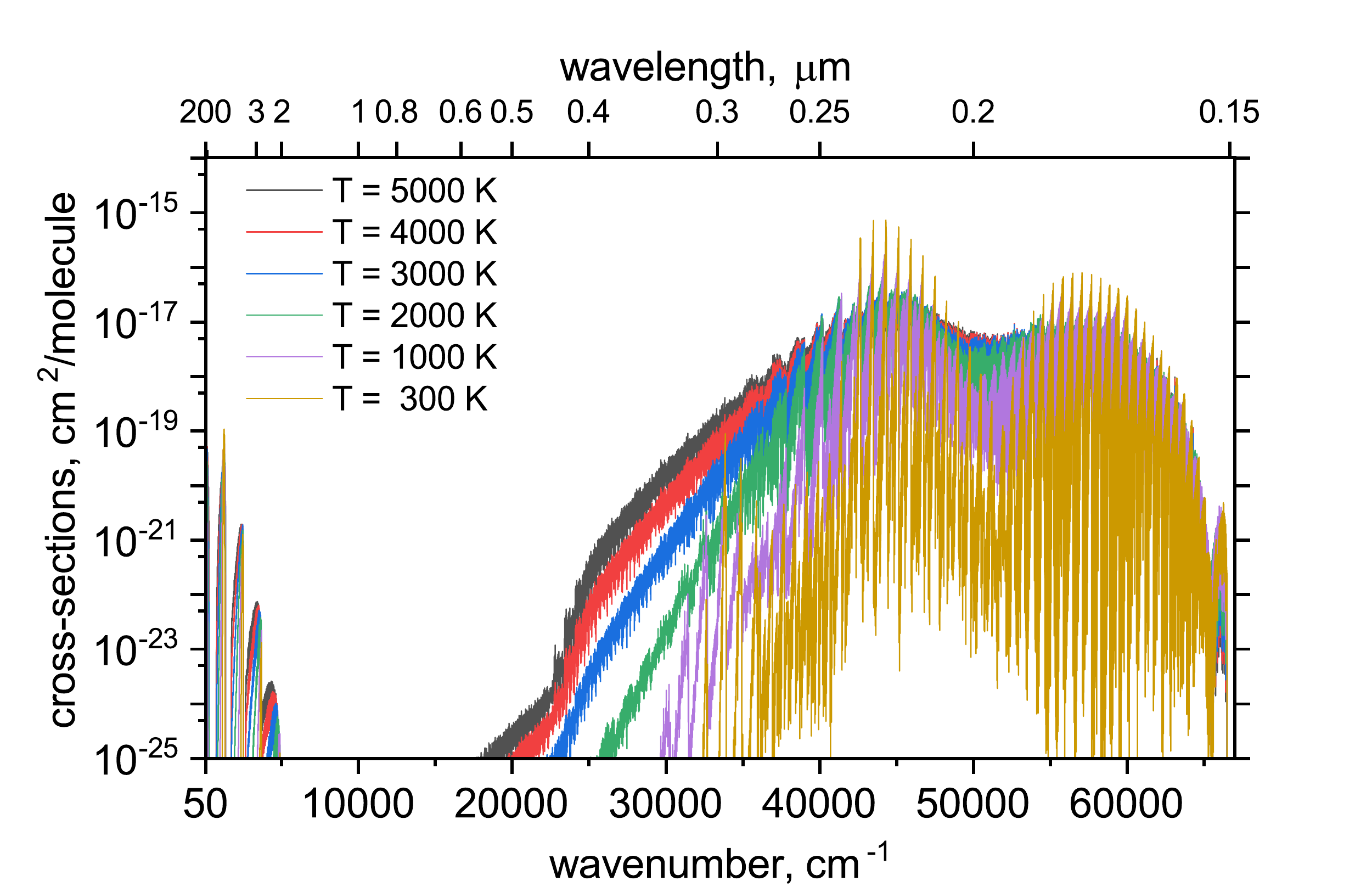}
    \caption{Temperature dependence of the SiO spectra using the \name\ line list. A Gaussian profile with half width at half maximum (HWHM) of 1~\cm\ was used.}
    \label{fig:spectra}
\end{figure}

Figure~\ref{fig:Bands} illustrates the three main bands of SiO compared against a spectrum generated from the \citet{11Kurucz.db} line list at 2000 K. We see increased deviation between the spectra with increased energy due to %\red{What is the cause of difference between us and Kurucz?}.
Figure \ref{fig:spectra} illustrates the evolution of the spectrum with increasing temperature.

%%%%%%%%%%%%%%%%%%%%%%%%%%%%%%%%%%%%%%%%%%%%%%%%%%%%%%%%%%%%%%%%%%%%%
%\begin{figure}
%\centering
%\includegraphics[width=0.80\textwidth]{SiO_vs_Kurucz_col%ourcode.png} %Model 52, Enermax 50, 000
%\caption{Simulated SiO absorption spectra at 2000 K. A %Gaussian profile of half width of half maximum (HWHM) of %1~\cm\ was used.}
%\label{f:XAE}
%\end{figure}
%%%%%%%%%%%%%%%%%%%%%%%%%%%%%%%%%%%%%%%%%%%%%%%%%%%%%%%%%%%%%%%%%%%%%

%\begin{figure}
%\centering
%\includegraphics[width=0.80\textwidth]{SiO_temp_depend.%png} %Model 52, Enermax 50, 000
%\caption{Simulated absorption spectra of SiO for a %range of temperatures. A Gaussian profile of half width %of half maximum (HWHM) of 1~\cm\ was used.}
%\label{f:temp}
%\end{figure}
%%%%%%%%%%%%%%%%%%%%%%%%%%%%%%%%%%%%%%%%%%%%%%%%%%%%%%%%%%%%%%%%%%%%%

\begin{figure}
    \centering
    \includegraphics[width=0.75\textwidth]{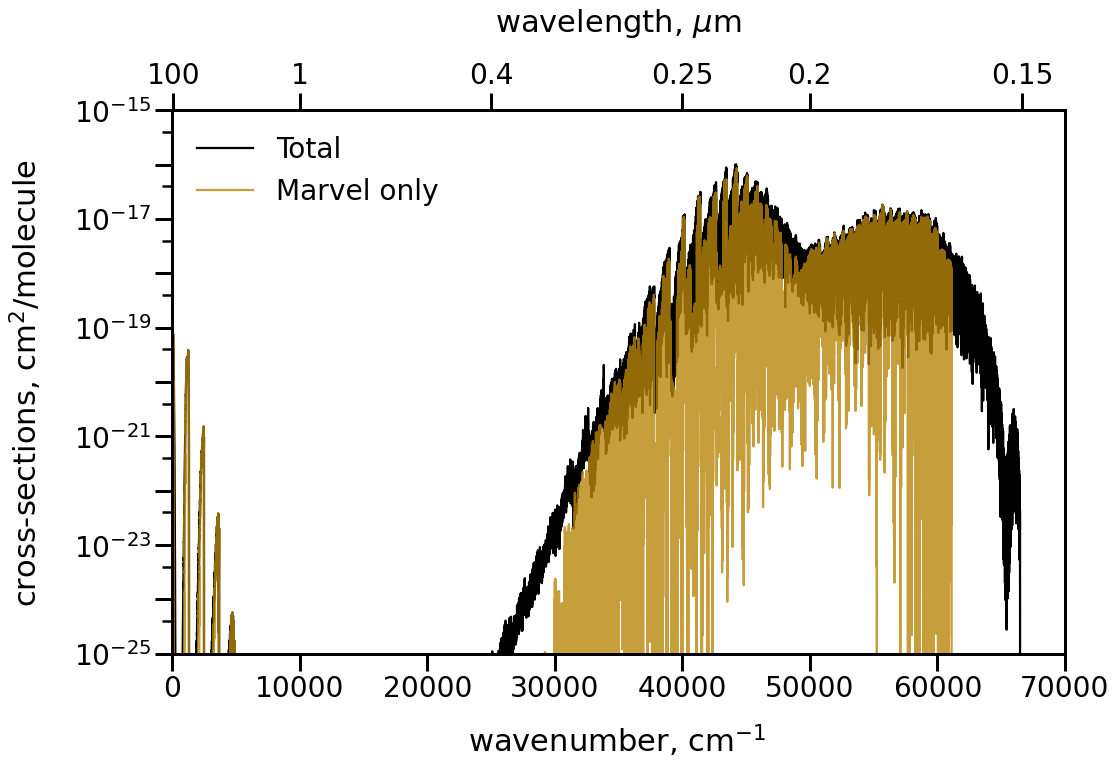}
    \caption{A SiO absorption spectrum at $T=$ 2000 K simulated using transitions between  MARVEL states only. A Gaussian profile with HWHM of 1~\cm was used. }
    \label{fig:spectrum-MARVEL}
\end{figure}

Figure~\ref{fig:spectrum-MARVEL} shows a spectrum of SiO computed using transitions between the `MARVELised' (i.e. more accurate) states only. It illustrates the completeness of the MARVEL part of the line list for high-resolution  spectroscopic applications, such  as high-dispersion spectroscopy  of exoplanets \citep{14Snellen}.

%Figure~\ref{fig:HeCoMo_Comparison} compares a simulated spectrum using the \name\ line list to the experimental spectrum by \citet{01HeCoMo.SiO}.

\begin{figure}
    \centering
    \includegraphics[width=0.5\textwidth]{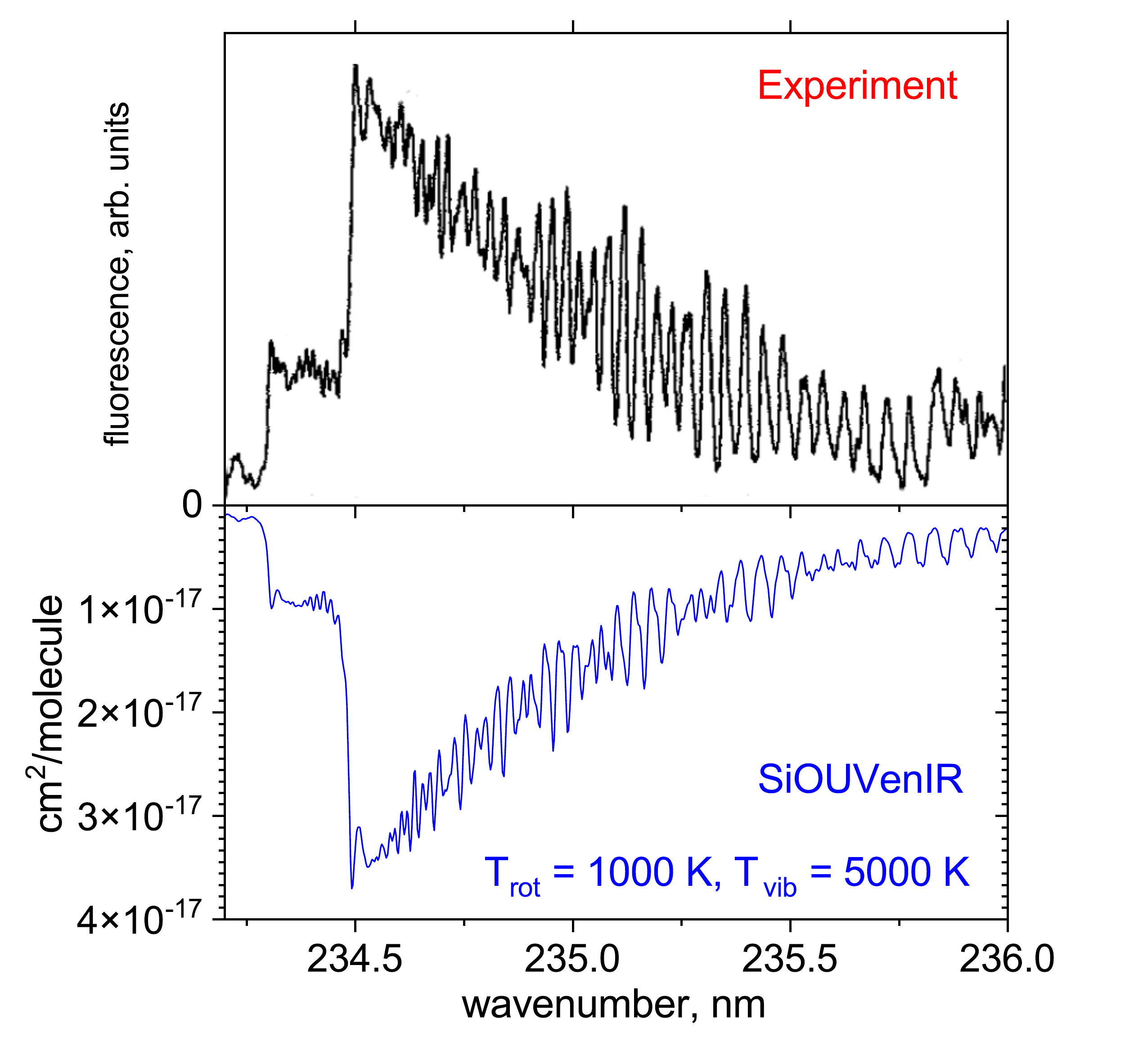}
    \includegraphics[width=0.45\textwidth]{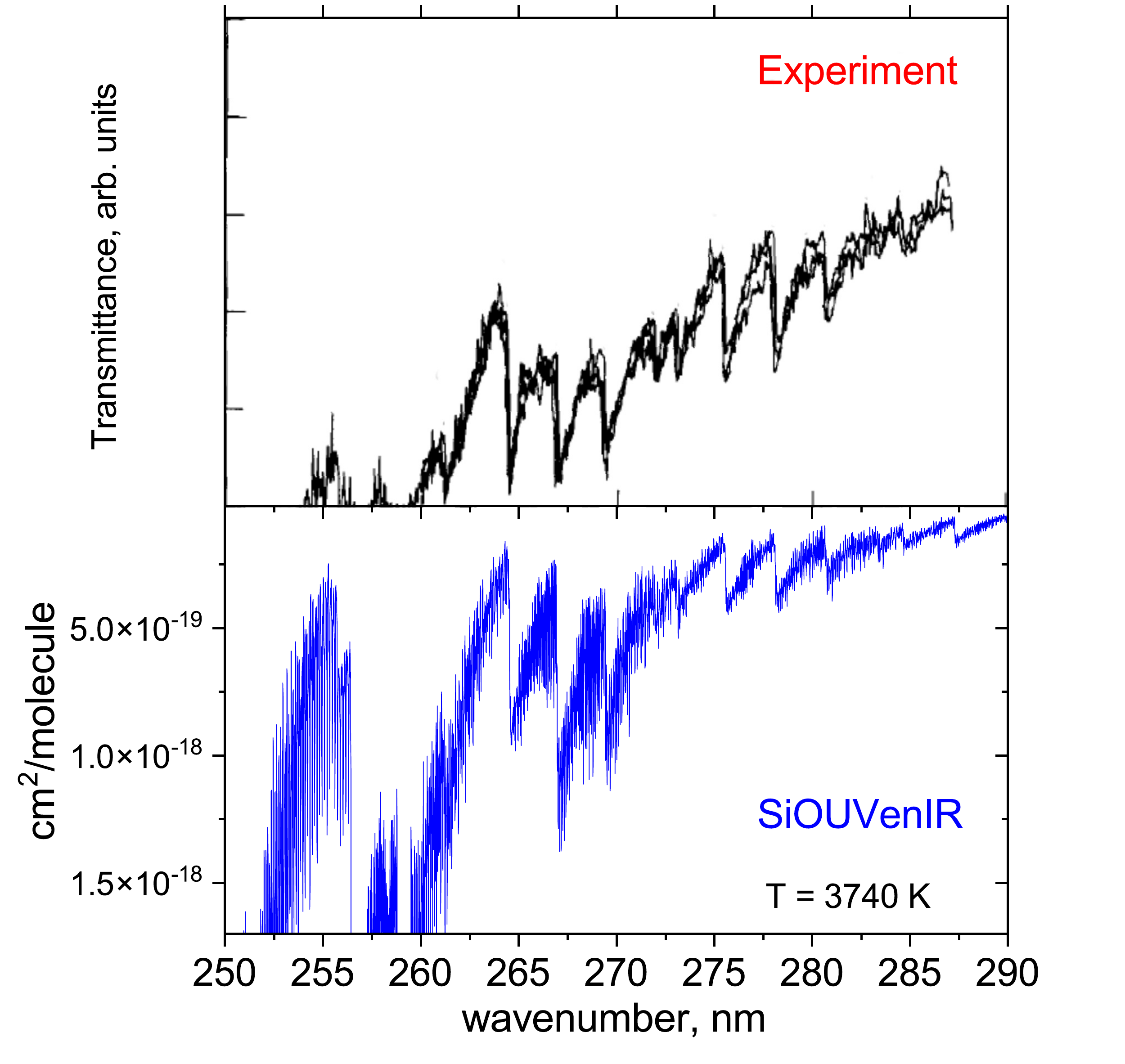}
    \caption{SiO $A-X$ band. Left display:  Absorption spectrum of SiO computed using $T_{\rm rot} = 1000$~K, $T_{\rm vib} = 5000$~K, using the Voigt line profile with  HWHM = 0.015~nm compared to a fluorescence spectrum  by \citet{03CoMoVi.SiO}; Right display:  SiO absorption spectrum  computed using $T$ = 3740~K and the Voigt line profile with using HWHM = 0.03~nm  compared to the shock-tube absorption spectrum by \citet{78PaArxx.SiO}. }
    \label{fig:03CoMoVi_Comparison}
\end{figure}

\begin{figure}
    \centering
    \includegraphics[width=0.45\textwidth]{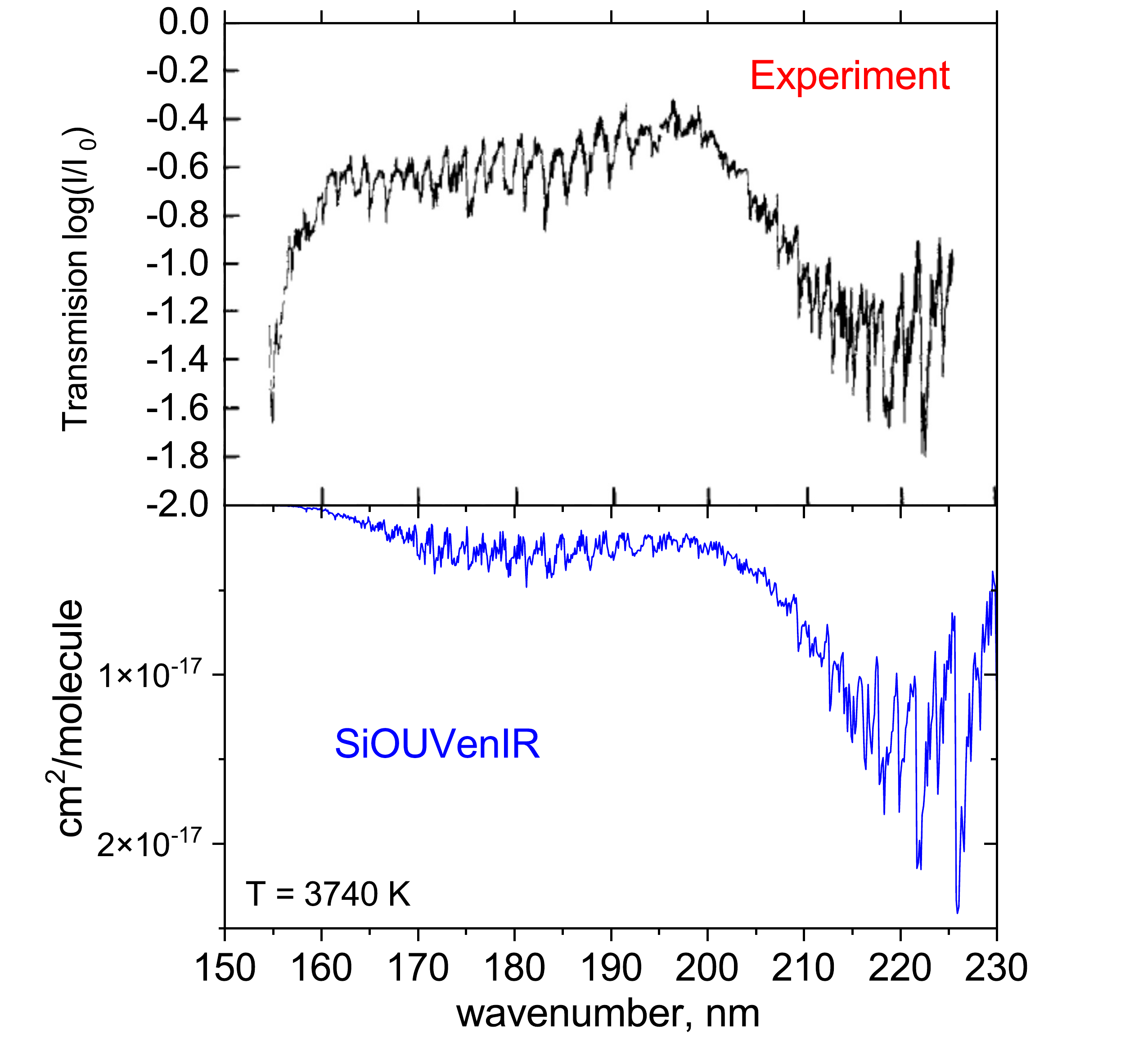}
    \caption{SiO $E-X$ band. An absorption spectrum of SiO  computed using $T$ = 3740~K and the  Voigt line profile compared to a shock-tube absorption spectrum by \citet{78Park.SiO} using HWHM = 0.04~nm.}
    \label{fig:78PaArxx}
\end{figure}

As an illustration of of the quality of the SiO line list in the UV, we compare theoretical SiO spectra with existing experimental spectroscopic studies of SiO from this region of two prominent bands, $A$--$X$ and $E$--$X$. Figure~\ref{fig:03CoMoVi_Comparison} compares  absorption spectra from the $A$--$X$ bands from two different regions, 234--236 nm and 250-290 nm to the  fluorescence spectrum by \citet{03CoMoVi.SiO} (Left) and shock-tube absorption spectrum  by \citet{78PaArxx.SiO} (Right), respectively. To reproduce the fluorescence spectrum at the 235 nm band, a non local thermodynamic equilibrium (non-LTE) population of $T_{\rm rot} = 1000$~K and $T_{\rm vib} = 5000$~K was used. Figure~\ref{fig:78PaArxx} provides an illustration for the  $E-X$ band by comparing the shock-tube absorption spectrum  by \citet{78Park.SiO} a \name\ synthetic spectrum  computed for  $T$ = 3740~K (LTE). All three cases demonstrate fairly good agreement with experiment.

\begin{figure}
    \centering
    \includegraphics[width=0.45\textwidth]{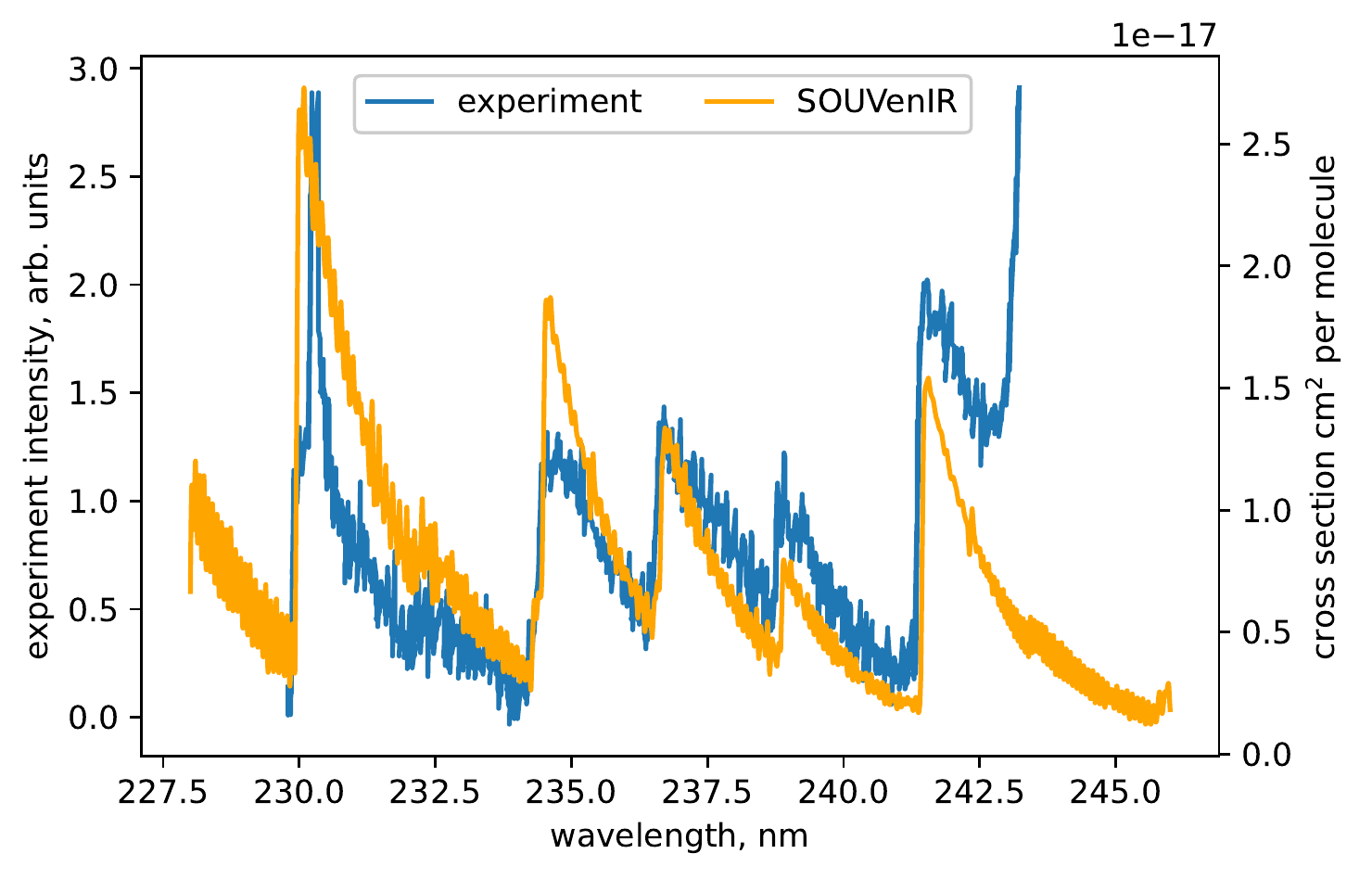}
    \caption{A comparison between cross sections generated from  \name~and the experimental data from \citet{01HeCoMo.SiO}. \name~cross sections are generated using Lorentzian line profiles with HWHM = 4.5~\cm, in simulated non-LTE with a rotational temperature of 3000~K and a vibrational temperature of 4000~K.}
    \label{fig:HeCoMo_Comparison}
\end{figure}

%\subsection{Stellar spectra comparisons}

\begin{figure}
    \centering
    \includegraphics[width=0.80\textwidth]{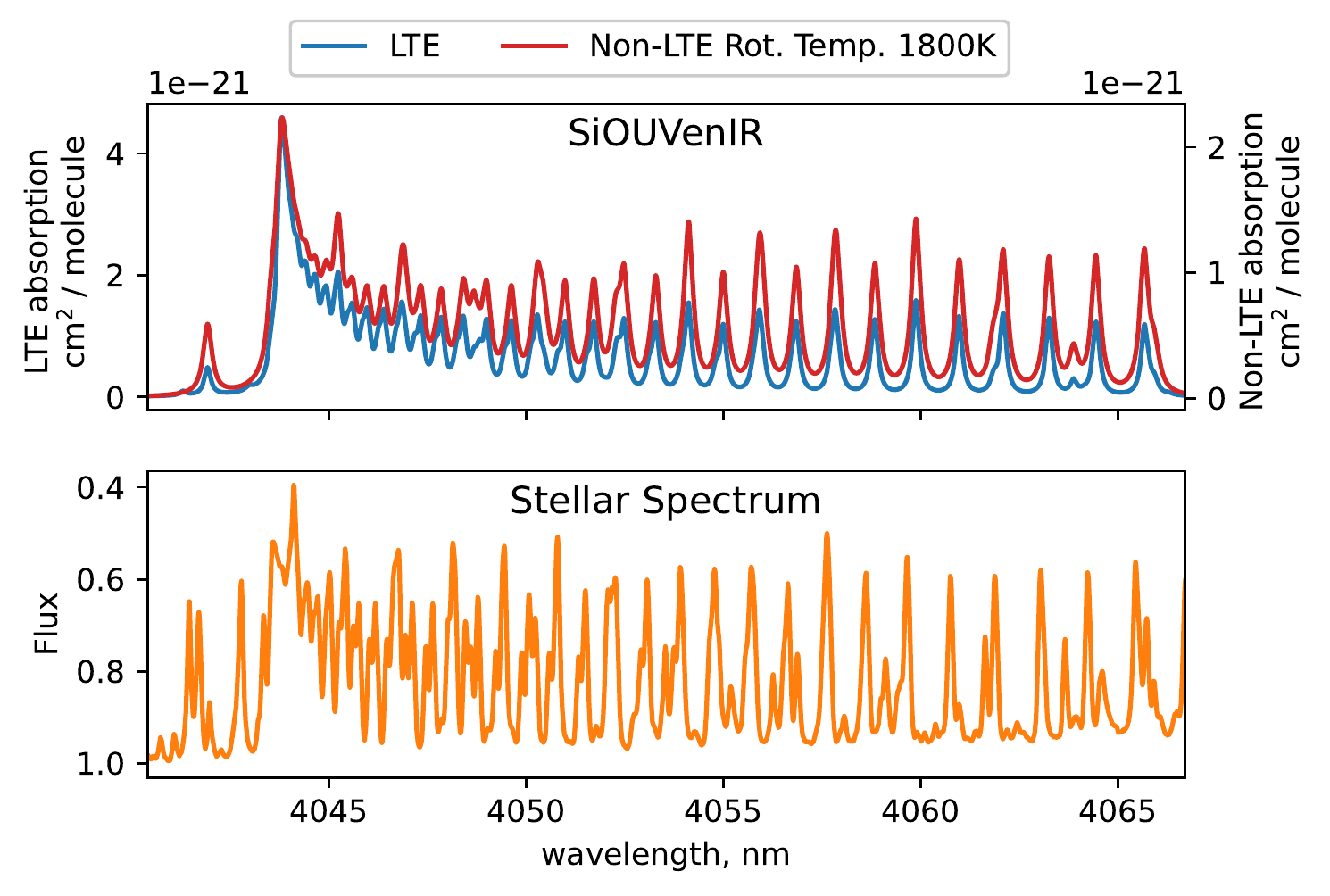}
    \caption{A comparison of the SiO spectrum (in air) using the \name\ line list with a section of the stellar spectrum of HD61913 from \citet{12LeSeUt.SiO} showing the 2-0 (middle) and 3-1 (left) IR band region.    A Lorentzian line profile with HWHM of 0.2~\cm\ was used. For the LTE spectrum, the equilibrium temperature of 3530~K was used. The non-LTE spectrum was computed for $T_{\rm rot} = 1800$~K and $T_{\rm vib} = 3530$~K.}
    \label{fig:HD61913_SiO_Cross_Sections}
\end{figure}

As an astrophysical application, Fig.~\ref{fig:HD61913_SiO_Cross_Sections} shows a simulation of the stellar spectrum of the star HD61913, taken from \citet{12LeSeUt.SiO}. This part of the spectrum features the 2--0 IR band, with the band head at 4044~nm. The small satellite bandhead belongs to the 3--1 hot band. The  absorption spectrum (air) of SiO  was generated from the \name\ line list using the Lorentzian line profile with a HWHM of 0.2~\cm\ with the Beer-Lambert law then applied  plotted next to the  HD61913 spectrum. In addition to the local thermodynamic equilibrium (LTE) spectrum calculated with the star's effective temperature of 3530~K, a non-LTE spectrum (calculated with a rotational temperature of 1800~K and a vibrational temperature of 3530~K) is also plotted. The non-LTE spectrum demonstrates improved similarity to the stellar spectrum, particularly with regard to the satellite 3--1 bandhead  next to the 2--0 band head.

%\subsection{Lifetimes}
%\label{subsec:lifetimes}

Lifetimes of SiO were reported in multiple theoretical \citep{03ChChD1.SiO,76OdElxx.SiO,79LaArxx.SiO,98DrSpEd.SiO,16Bauschlicher.SiO,19FeZhxx.SiO} and experimental \citep{72SmLixx.SiO, 73ElSmxx.SiO} works. Table~\ref{t:lifetimes} shows a comparison of these works with  the calculated lifetimes for the \A\ and \E states. In our work, the output from \textsc{Duo}  was used as the input for the \textsc{ExoCross} program \citep{jt708} to calculate lifetimes, see also \cite{jt624}. It should be noted that the theoretical  lifetimes show slow dependence on $J$ and tend to monotonically increase with $v$ up to about 100 ns for $v=40$. This means that an
observed lifetime for the state will be temperature dependent.

For the \A\  state at $v'=0, J'=0$ our lifetime value is 28.5 ns, which is in close agreement with some previous theoretical works \citep{16Bauschlicher.SiO,19FeZhxx.SiO,03ChChD1.SiO}, but is twice the calculated lifetime predicted by \citet{79LaArxx.SiO} and \citet{98DrSpEd.SiO}, which seem to be closer to the experimental value of \citet{72SmLixx.SiO}. While it is hard for us to reconcile these large differences, as we use a very high accuracy \ai\ model for TDMCs, we attribute the difference with the experimental work \citet{72SmLixx.SiO} to be due to the lower perturbing states.

For the \E\ state at $v'=0, J'=0$ our value is 11.23 ns, which is in close agreement with  some very recent calculations  \citep{16Bauschlicher.SiO, 19FeZhxx.SiO}  and experiment \citep{73ElSmxx.SiO}.

%%%%%%%%%%%%%%%%%%%%%%%%%%%%%%%%%%%%%%%%%%%%%%%%%%%%%%%%%%%%%%%%%%%%%%%%%%%%%%%%%%%%%%%%%%%%%%%%%%%%%%%%%%%%%%%%%%%%%%%%%%%%%%%%%%%%%%%%%%%%
\begin{table}
\centering
\caption{Comparison of the experimental and calculated lifetimes, $\tau$ (ns), for the $A ^{1}\Pi$ and $E ^{1}\Sigma^{+}$ vibrational ground states. `This work' values are quoted for $v'=0, J'=0$. }
%\red{ Ahmad, Are all decimal places as in the original publications, e.g. 6.00 ns? (Ahmad: Yes)}

%\tt
\label{t:lifetimes}
\centering
\begin{tabular}{rrrcrr}
\hline\hline
\multicolumn{1}{c}{$A ^{1}\Pi$}	&	\multicolumn{1}{c}{source} &	& \multicolumn{1}{c}{$E ^{1}\Sigma^{+}$}	&\multicolumn{1}{c}{source} & \\
\hline
 {28.5} & This work             & Cal  &  {11.23} & This work & Cal \\
 {9.6} & \citet{72SmLixx.SiO}  & Exp  &  {10.5} & \citet{73ElSmxx.SiO} & Exp  \\
        28.9 & \citet{03ChChD1.SiO}  & Cal  &       7.10  & \citet{03ChChD1.SiO} & Cal \\
        49.5 & \citet{76OdElxx.SiO}  & Cal  &             &  \\
        16.6 & \citet{79LaArxx.SiO}  & Cal  &      6.80   & \citet{79LaArxx.SiO} & Cal \\
        12.5 & \citet{98DrSpEd.SiO}  & Cal  &      6.00   & \citet{98DrSpEd.SiO} & Cal \\
        29.3 & \citet{16Bauschlicher.SiO}  & Cal  &     11.0    & \citet{16Bauschlicher.SiO} & Cal  \\
        32.34 & \citet{19FeZhxx.SiO} & Cal  &     10.58   & \citet{19FeZhxx.SiO} & Cal \\
\hline\hline
\end{tabular}
\begin{flushleft}
\noindent \vspace*{0.01truecm}

\end{flushleft}
\end{table}

\label{sec:stellar}

\section{Conclusion}
\label{sec:conc}

A new IR and UV line list for SiO is presented. The \name\ line list supersedes the original SiO ExoMol line list from \citet{jt563}.  \name\ is available from \url{www.exomol.com} and the CDS database, via
ftp://cdsarc.u-strasbg.fr/pub/cats/J/MNRAS/, or
http://cdsarc.u-strasbg.fr/viz-bin/qcat?J/MNRAS/

As part of the line list construction, a MARVEL analysis for \sio\ was performed. All experimental line positions  from the literature (to the best of our knowledge) covering the IR and UV regions ($X$--$X$, $A$--$X$ and $E$--$X$ systems) were collected and processed to generate a comprehensive set of empirical energies of \sio. Forbidden experimental transitions connecting the $X$ state with the dark electronic states of the $C$, $D$, $e$, and $d$ system were assigned using our fully coupled description of the rovibronic system.
An accurate spectroscopic model for SiO was built: \ai\ TDMC and empirically refined PECs, SOCs, EAMCs. This allowed us to make full line assignments
to transitions associated with perturbing states.

The line list was MARVELised, where the theoretical energies are replaced  with the MARVEL  values (where available). The  line list provides uncertainties of the rovibronic states in order to help in high-resolution applications.  Comparisons of simulated UV spectra show close agreement with experiment, similarly with computed lifetimes.

This \name\ linelist was largely constructed through two online ``Hackathon" days during 2020  via the online video-conferencing Zoom platform. The team was split into ``Breakout rooms" which each had a particular task (\textit{Ab initio}, MARVEL, fitting, etc.). These tasks were largely independent for initial iterations of this new linelist.  The online Hackathon days had the advantage of including co-authors who would not normally be able to participate in ``in-person" ``Hackathons".

%\green{From JT:  OK so I went back to Emma's report to get residues with vibrational excitation. You are right for v up to 6 the results are excellent. Please can you amend the errors as follows: \\ For v=0, use your J-dependent formula For v = 1 to 6 simply add 0.04 cm-1 as a constant to the uncertainty. \\ You seem to go up to v = 7. This is (probably) less accurate eg 0.1 to 0.2 cm-1. It would probably be best just to drop it if we do not lose anything. Perhaps you could supply 2 versions: \\ a. up to v=6 only b. with v=7 and use 0.2 cm-1 as the base vibrational uncertainty.}

\section*{Acknowledgments}

This work was supported by the STFC Projects No. ST/R000476/1, ST/S506497/1 and ST/P006736/1 and
by the European Research Council (ERC) under the European Union’s Horizon 2020 research and innovation programme through Advance Grant number 883830. The authors acknowledge the use of the UCL Legion High Performance Computing Facility (Legion@UCL) and associated support services in the completion of this work, along with the Cambridge Service for Data Driven Discovery (CSD3), part of which is operated by the University of Cambridge Research Computing on behalf of the STFC DiRAC HPC Facility (www.dirac.ac.uk). The DiRAC component of CSD3 was funded by BEIS capital funding via STFC capital grants ST/P002307/1 and ST/R002452/1 and STFC operations grant ST/R00689X/1. DiRAC is part of the National e-Infrastructure. This research was undertaken with the assistance of resources from Supercomputing Wales and the
Australian National Computational Infrastructure (NCI Australia), a NCRIS enabled capability supported by the Australian Government. Additional supported was provided by UK research councils EPSRC, under grant EP/N509577/1. We would like to thank the UK Natural Environment Research Council (NERC) for funding through grant NE/T000767/1. We also want to thank Moscow Witte University for sponsoring the fellowship enabling this research.

%\red{More thanks to STFC (Wilf, Sam)}
%\red{just to note have added in Supercomputing Wales - hope this okay (mg).}

\section{Data Availability} %Template incase we want to include!
The data underlying this article are available in the article and in its online supplementary material.

\section*{Supporting Information}
The MARVEL input transitions file and output energy levels file, the \textsc{Duo} input file, which contains all the potential energy, dipole moment and coupling curves of SiO used in this work, and the  temperature-dependent partition function of \sio\ up to 10000~K are given as supplementary only data to this
article. The full \name\ line list is available
from \url{www.exomol.com} and the CDS database, via
ftp://cdsarc.u-strasbg.fr/pub/cats/J/MNRAS/, or
http://cdsarc.u-strasbg.fr/viz-bin/qcat?J/MNRAS/.

\bibliographystyle{mnras}
%\bibliographystyle{mn2e}

%in order to connect to a  bibliography file from github
%Upload -> from external URL -> add, e.g.
%add a https://raw.githubusercontent.com/ExoMol/bib/master/exomol/comets.bib
%\bibliography{journals_astro,jtj,SiO,methods,abinitio,programs,Books,linelists,partition,planets}

%\bibliography{journals_astro,jtj,SiO,linelists,methods,exoplanets,abinitio,programs,CN,Books,MARVEL}

%\label{lastpage}

\end{document}